\documentclass[traditabstract]{aa}

\usepackage{graphicx}
\usepackage{natbib}
\usepackage{multirow}
\usepackage{xspace}
\usepackage{amsmath}
\usepackage{url}
\usepackage[varg]{txfonts}
\usepackage{xcolor}
\newcommand{\xmm}{\mbox{XMM-Newton}\xspace}
\newcommand{\integral}{INTEGRAL\xspace}
\newcommand{\suzaku}{Suzaku\xspace} 
\newcommand{\rxte}{RXTE\xspace}
\newcommand{\chandra}{Chandra\xspace}
\newcommand{\nustar}{NuSTAR\xspace}

\newcommand{\pn}{{\mbox{EPIC-pn}}\xspace}

\newcommand{\pca}{PCA\xspace}
\newcommand{\hexte}{HEXTE\xspace}
 
\newcommand{\isgri}{ISGRI\xspace}
\newcommand{\cyg}{{\mbox{Cyg~X-1}}\xspace}

\bibpunct{(}{)}{;}{a}{}{,} 

\graphicspath{{./}{figs/}}
\begin{document}

\title{Revealing the broad iron K$\alpha$ line in \cyg through
  simultaneous XMM-Newton, RXTE, \& INTEGRAL observations}
\author{Refiz Duro\inst{1,2} 
  \and Thomas Dauser\inst{1} 
  \and Victoria Grinberg\inst{3} 
  \and Ivica Mi\v skovi\v{c}ov\'a\inst{1} 
  \and J\'er\^ome Rodriguez\inst{4} 
  \and \mbox{John Tomsick}\inst{5} 
  \and \mbox{Manfred Hanke}\inst{1}
  \and Katja Pottschmidt\inst{6,7} 
  \and Michael A. Nowak\inst{3}
  \and Sonja Kreykenbohm\inst{1,8}
  \and Marion Cadolle Bel\inst{9}
  \and \mbox{Arash Bodaghee}\inst{5,10} 
  \and Anne Lohfink\inst{11}
  \and Christopher S. Reynolds\inst{12}
  \and Eckhard Kendziorra\inst{8}\thanks{deceased 2 September 2015}
  \and Marcus G. F. Kirsch\inst{13}
  \and \mbox{R\"udiger Staubert}\inst{8}
  \and  \mbox{J\"orn Wilms\inst{1}} } 
\institute{Dr.\ Karl Remeis-Sternwarte and Erlangen
  Centre for Astroparticle Physics, Friedrich-Alexander-Universit\"at
  Erlangen-N\"urnberg, Sternwartstra\ss{}e~7, 96049 Bamberg, Germany
  \and
  AIT Austrian Institute of Technology GmbH, Donau-City-Stra\ss{e}~1,
  1220 Vienna, Austria
  \and 
  Kavli Institute for Astrophysics,
  Massachusetts Institute of Technology, Cambridge, MA 02139, USA 
  \and 
  Laboratoire AIM, CEA/IRFU-Universit\'e Paris Diderot-CNRS/INSU,
  CEA DSM/IRFU/SAp, Centre de Saclay, F-91191 Gif-sur-Yvette, France  
  \and 
  Space Sciences Laboratory, University of California, 7 Gauss
  Way, Berkeley, CA 94720, USA 
  \and 
  CRESST and NASA Goddard Space Flight Center, Astrophysics Science
  Division, Code 661, Greenbelt, MD 20771, USA 
  \and
  Center for Space Science and Technology,
  University of Maryland Baltimore County, 1000 Hilltop Circle,
  Baltimore, MD 21250, USA 
  \and 
 Institut f\"ur Astronomie und Astrophysik, Eberhard Karls
 Universit\"at T\"ubingen, Sand 1, 72074 T\"ubingen, Germany 
 \and
 Max Planck Computing and Data Facility, Gie\ss{}enbachstra\ss{}e~2, 85748 
  Garching, Germany 
  \and
  Department of Chemistry, Physics, and Astronomy, Georgia College, CBX
  082, Milledgeville, GA 31061, USA 
  \and
  Institute of Astronomy, Madingley Road, Cambridge CB3 0HA, United Kingdom
  \and 
  Department of Astronomy and Maryland Astronomy Center for Theory and
  Computation, University of Maryland, College Park, MD 20742, USA 
 \and 
  European Space Agency, European Space Operations Centre,
  Robert-Bosch-Stra\ss{}e 5, 64293 Darmstadt, Germany }

\date{Accepted 20/02/2016}
 
\abstract{We report on the analysis of the broad Fe K$\alpha$ line
  feature of \cyg in the spectra of four simultaneous hard
  intermediate state observations made with the X-ray Multiple Mirror
  mission (\xmm), the Rossi X-ray Timing Explorer (\rxte), and the
  International Gamma-Ray Astrophysics Laboratory (\integral). The
  high quality of the \xmm data taken in the Modified Timing Mode of
  the \pn camera provides a great opportunity to investigate the
  broadened Fe K$\alpha$ reflection line at 6.4\,keV with a very high
  signal to noise ratio. The 4--500\,keV energy range is used to
  constrain the underlying continuum and the reflection at higher
  energies. We first investigate the data by applying a
  phenomenological model that consists of the sum of an exponentially
  cutoff power law and relativistically smeared reflection.
  Additionally, we apply a more physical approach and model the
  irradiation of the accretion disk directly from the lamp post
  geometry. All four observations show consistent values for the black
  hole parameters with a spin of $a\sim 0.9$, in agreement with recent
  measurements from reflection and disk continuum fitting. The
  inclination is found to be $i\sim30\degr$, consistent with the
  orbital inclination and different from inclination measurements made
  during the soft state, which show a higher inclination. We speculate
  that the difference between the inclination measurements is due to
  changes in the inner region of the accretion disk.}
 
\authorrunning {Duro et al.} \titlerunning{Broad Fe K$\alpha$ line of
  \cyg}
\maketitle
%

\section{Introduction}\label{sec:intro}

The high mass X-ray binary system \cyg
\citep{bowyer:1965,murdin:1971} consists of a $14.8\pm 1.0\,M_\odot$
black hole and the $19.2\pm1.9\,M_\odot$ OB giant star HDE~226868
\citep[][but see also \citealt{ziolkowski:2014}]{orosz:2011}. The
system is at a distance of $d\sim1.86$\,kpc
\citep{reid:2011,xiang:2011}. The orbital period of the system is
5.6\,d \citep{webster:1972,bolton:1972} at an inclination of
$i=27\fdg1\pm0\fdg8$ \citep{orosz:2011}. Recently, several groups
reported a very high spin of $a\sim0.9$ for \cyg
\citep[][]{duro:2011,gou:2011,fabian:2012}. The donor star HDE~226868
is close to filling its Roche lobe \citep{gies:1986}, creating a
focused wind towards the black hole \citep{friend:1982}, where the
wind material forms an accretion disk. The observed X-ray flux from
the disk is orbital-phase dependent, since the wind material can
absorb a significant amount of the flux, resulting in spectral
absorption and, in the most extreme cases, dips in the lightcurves of
the observed object \citep[see,
  e.g.,][]{Grinberg_2015a,miskovicova:2013,hanke:2009,diaz:2006,balucinska:2000}.

Historically, \cyg has spent most of its time in the low luminosity
``hard state'' \citep{wilms:2006,grinberg:2013}, in which the X-ray
spectrum above $\sim$2\,keV is well described by an exponentially
cutoff power-law with a photon index of $\Gamma\sim1.7$ and an
exponential folding energy of $E_\mathrm{fold}\sim150\,\mathrm{keV}$
\citep{sunyaev:79a,wilms:2006,nowak:2011}, and a very weak disk
component. The power-law emission originates in a corona or a jet
\citep[e.g.,][]{nowak:2011}. Radio emission is detected during the
hard state and is attributed to synchrotron emission from electrons in
a relativistic outflow \citep{stirling:2001,malzac:2009}. Synchrotron
radiation is also the likely cause for the hard tail seen at
$>$400\,keV \citep{mcconnell:94a,mcconnell:2000,cadollebel:2006},
which is strongly polarized \citep{laurent:2011,jourdain:2012}. In the
``soft state'' the spectrum can be described by thermal emission from
a 0.5--0.6\,keV standard accretion disk \citep[see,
  e.g.,][]{tomsick:2014}, a steep power-law photon index, and no high
energy cutoff. Occasional transitions and failed transitions from one
to the other state define a ``hard (or soft) intermediate state'',
that is characterized by distinct spectral and variability properties
\citep{pottschmidt:2003,grinberg:2014}.

The behavior of \cyg can be explained by the underlying geometry of an
accreting black hole. In the standard view, the black hole is
surrounded by an accretion disk that is illuminated by photons from a
corona or the base of a jet, which is located close to the black hole.
At the inner edge of the accretion disk, the temperature is typically
a few 100\,eV. A fraction of the disk's thermal emission is
intercepted by the thermal (or hybrid) hot gas in the vicinity of the
disk and the black hole (see \citealt{reynolds:2003} and references
therein for a more detailed discussion of potential accretion
geometries). The intercepted soft X-ray photons are Compton
up-scattered by the hot electrons and produce a hard spectral
component. The close correlation between the X-ray and the radio
emission on timescales of days and weeks suggests a close coupling
between them \citep{hannikainen:1998,corbel:2000,gleissner:2004}. Such
observations have resulted in an alternative model for the emission
process, in which a jet may play the role of the hard X-ray source
\citep{martocchia:1996,markoff:2005}. In this model, the radio
emission is produced by synchrotron emission from the relativistic
electrons in the jet, while the hard X-rays are due to a combination
of this emission and Synchrotron self-Comptonization emission from the
base of the jet. As shown, e.g., by \citet{nowak:2011}, both types of
models can describe the observed data equally well.

Common to both geometrical models is the fact that a fraction of the
hard X-ray photons is reflected by the accretion disk, producing
fluorescent line emission, which is then broadended by relativstic
effects \citep{fabian:1989, laor:1991}.

The Fe K$\alpha$ line is the strongest of such fluorescent emission
lines and is observed in many galactic black hole systems \citep[][and
  references therein]{nowak:2002,reis:2009,duro:2011,gou:2014} and
active galactic nuclei \citep[AGN,][and references
  therein]{tanaka:1995,fabian:2009,dauser:2012}, see
\citet{reynolds:2003} and \citet{miller:2007} for reviews. Since it
originates from the reflection of hard X-ray photons at the innermost
regions of the accretion disk, the line bears the imprint of Doppler
effects and relativistic and gravitational physics close to the black
hole. These effects distort the intrinsically narrow shape of the
line, leading to a broad, redshifted, and skew-symmetric spectral
profile \citep{fabian:1989,laor:1991}.

\begin{figure*}
  \centering
  \resizebox{\hsize}{!}{\includegraphics{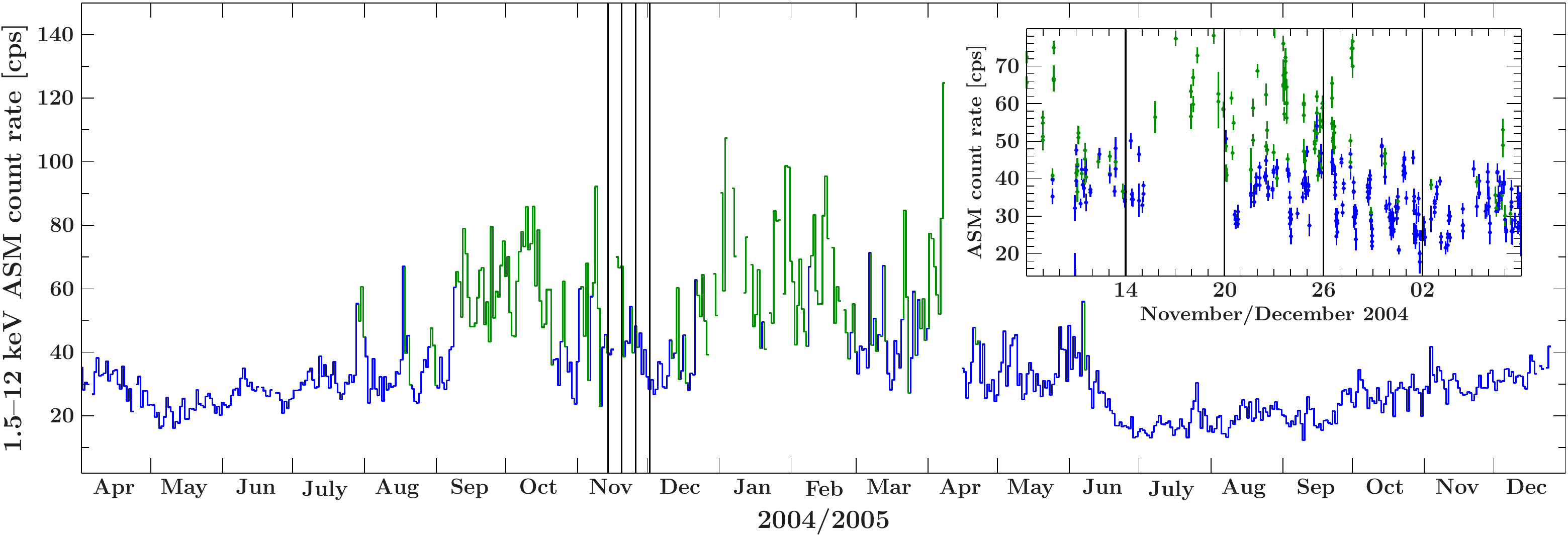}}
  \caption{\rxte All Sky Monitor lightcurve of Cyg~X-1 rebinned to a
    resolution of 1\,d for the two year period surrounding the
    observations. The times of our four analyzed observations are
    marked with the vertical lines. Color coding corresponds to
    spectral state classification as defined by \citet{grinberg:2013}:
    the blue color represents hard state, while the green color
    represents the intermediate state data. The inset is zoomed
    in on the four observations, with a resolution of one data point
    per spacecraft orbit.}
  \label{fig:asm}
\end{figure*}

We can identify the inner disk radius, $R_\mathrm{in}$, with the
innermost stable circular orbit (ISCO). This idea is supported, for
example, by the work of \citet{miller:2012}, which suggests that the
disk remains close to the ISCO even during the hard state of \cyg
\citep[see also][]{young:1998, dovciak:2004, reynolds:2008}. This
assumption allows us to measure directly the spin of the black hole.
Namely, the radius $R_\mathrm{ISCO}$ depends on the dimensionless spin
parameter, $a$, of the black hole, where $-0.998 \leq a \leq +0.998$
\citep{thorne:1974}, which translates to $9 r_\mathrm{g} \geq
R_\mathrm{ISCO} \geq 1.24 r_\mathrm{g}$, where $r_\mathrm{g}=GM/c^{2}$
is the gravitational radius\footnote{Negative spin values mean that
  the angular momenta of the black hole and the disk are
  anti-parallel.}. A larger spin value means that the inner edge of
the accretion disk is closer to the black hole, so that an increasing
number of photons originate from the innermost region and alters the
line shape. These additional photons undergo the strongest
gravitational redshift because of their origin close to the black
hole, leading to a stronger red wing in the line profile. Examining
this apparent ``broadening'' of the line we can measure the spin. In
addition, assuming that the spin of the black hole and the angular
momentum of the inner disk are aligned, the line profile can also be
used to measure the inclination of the system, $i$. Finally, the line
profile also depends on the ``emissivity profile'', which
characterizes the intensity irradiating the accretion disk as a
function of the radial distance, $r$, from the black hole and is
generally parametrized as a power-law, $r^{-\epsilon}$.

The Fe K$\alpha$ method is especially well suited for spin
measurements of compact accreting objects since it is completely
independent of the mass and distance measurements of the source and
only requires the presence of an accretion disk down to the innermost
stable circular orbit. For objects such as Cyg~X-1, for which the mass
and the distance are known, the Fe K$\alpha$ method is complementary
to the continuum-fitting method in which the accurate measurement of
these quantities is essential \citep{zhang:1997,li:2005}. In addition,
the method does not depend on the X-ray spectral state of the source,
in contrast to the continuum fitting method, which can only be used in
the soft state, when the disk emission is prominent. Both techniques
for determining the black hole spin have been applied to data of \cyg,
with the latest results picturing the source as a highly rotating one
\citep{duro:2011,gou:2011,fabian:2012,gou:2014,tomsick:2014}.

In this paper we study very high signal to noise observations of the
broad line feature in \cyg obtained in simultaneous \xmm, \integral,
and \rxte observations. In Sect.~\ref{sec:observations}, we describe
the reduction and the calibration of the \xmm Modified Timing Mode
data (Sect.~\ref{sec:pn}), which is crucial for accurate iron line
modeling, and the reduction of the \rxte (Sect.~\ref{sec:rxte}) and
\integral data (Sect.~\ref{sec:integral}). The analysis of the data
for different system geometries is presented in
Sect.~\ref{sec:relline}, where we use a phenomenological description
based on a cutoff power-law continuum and relativistically broadened
reflection. We first assume an emissivity profile with a power-law
shape, and then also successfully apply the more physically motivated
lamp post geometry. We find that \cyg is rapidly spinning. We discuss
these findings in Sect.~\ref{sec:discuss}, and summarize our results
in Sect.~\ref{sec:summary}.

\section{Data Reduction}\label{sec:observations}
 
\subsection{\xmm\,\pn data}
\label{sec:pn}

In order to determine the parameters of the black hole, we need to
measure a line profile with a very high signal to noise ratio,
$\mathrm{S}/\mathrm{N}$ in as little time as possible. This approach
allows us to reduce as much as possible the influence of profile
variations due to the source's spectral variability. The instrument
suited best for such requirements is \xmm's \pn camera
\citep{strueder:2001}. This detector has two observing modes suited
for observations of bright sources: the Burst Mode with only 3\%
live-time that results in low $\mathrm{S}/\mathrm{N}$, and the Timing
Mode with 99.5\% live-time.

In order not to distract from the science, the details of the modes
and their limitations are discussed in more detail in the appendices.
Here we only give a brief summary. While the Timing Mode yields
observations with low pile-up for fluxes up to $\sim$150\,mCrab, in
practice the telemetry allocated to the \pn instrument limits the
observations to only $\sim$100\,mCrab. This is much less than the
typical flux of \cyg, which can be as bright as $\sim$300\,mCrab. In
order to observe the source and acquire high $\mathrm{S}/\mathrm{N}$
data, we implemented a modification to the standard Timing Mode, the
so-called \textsl{Modified Timing Mode} \citep{kendziorra:2004}. The
idea of this mode is to reduce telemetry drop outs as much as
possible. In order to do so, all available EPIC telemetry is made
available to the EPIC-pn camera by switching off the EPIC-MOS cameras.
This can be done without loss of information, since MOS data are piled
up for sources above $\sim$35\,mCrab. Even with the MOS cameras
switched off, the telemetry needs of the Timing Mode are such that
telemetry drop outs caused by buffer overflows onboard the
\xmm-spacecraft would still occur. In order to avoid these drop outs,
we reduce the telemetry needs further by only transmitting those
events which are necessary for our prime science, the study of the Fe
K$\alpha$ line. This is done by increasing the lower energy threshold
of the EPIC-pn camera, i.e., the energy, above which data are
telemetered to ground from 0.2\,keV to 2.8\,keV. In combination, the
event rates decrease well below the new telemetry bandwidth of
$\sim$1000\,cps, so that \cyg becomes observable without telemetry
gaps and with manageable pile up. A disadvantage of the Modified
Timing Mode is that it requires a recalibration of the instrument.
Fortunately, this can be done based solely on available Timing Mode
data. We give a full description of the data mode, the new calibration
and its verification in Appendix~\ref{app:A}.

\begin{table}
  \centering
  \caption{Exposure times in ks for \xmm, \rxte, and \integral data.}
  \label{tab:exposure} 
    \begin{tabular}{c@{\hspace*{1mm}}ccccc}
      \hline
      Obs & Date & XMM ID & \pn & \pca \&  & \isgri \\
      & 2004 & & & \hexte &  \\
      \hline
      1 & Nov 14 & 0202760201 & 17.4 & $\phantom{0}$6.6  & 53.3 \\
      2 & Nov 20 & 0202760301 & 17.4 & $\phantom{0}$5.9  & 31.1 \\
      3 & Nov 26 & 0202760401 & 19.8 & 10.4 & 38.7 \\
      4 & Dec 02 & 0202760501 & $\phantom{0}$9.7 & $\phantom{0}$5.2  & 34.3 \\
      \hline
    \end{tabular}
  \tablefoot{    
    The exposure times provided are for each
      instrument.
  }
\end{table}

\subsection{Observing Cyg~X-1 with the Modified Timing
  Mode}\label{sec:source_state}

In 2004 November and December, using the Modified Timing Mode we
performed four \xmm observations of \cyg with exposure times of
$\sim$20\,ks for observations Obs1, Obs2, Obs3, and $\sim$10\,ks for
observation Obs4 (see Table \ref{tab:exposure} for the observation
logs for all instruments used). Figure~\ref{fig:asm} shows the X-ray
lightcurve of Cyg~X-1 as measured with RXTE's All Sky monitor for the
2004 April to 2005 December period. Shortly before the observations
Cyg~X-1 was in a prolonged hard state of low luminosity. This phase
was followed by an episode during which the 2--10\,keV flux varied and
during which our observations happened. We apply the state
classification scheme of \citet{grinberg:2013} for a more detailed
analysis of the source state. Only observations 1 and 3 have strictly
simultaneous ASM measurements. In both cases the position on the ASM
hardness-intensity diagram indicates an intermediate state. ASM
measurements taken within 6\,hours around observations~2 and~4 are not
conclusive and show the source either in the hard or in the
intermediate state. In all cases the observations are close to the cut
between the hard and intermediate state on the ASM
hardness-intensity-diagram \citep[see][Fig. 5]{grinberg:2013}, where
caution in state classification is advised. Given the high variability
of the intermediate state and the overall behavior of the source,
however, it is likely that \cyg was in the hard-intermediate state for
all four observations analyzed here.
   
\begin{figure}
  \centering
  \resizebox{\hsize}{!}{\includegraphics{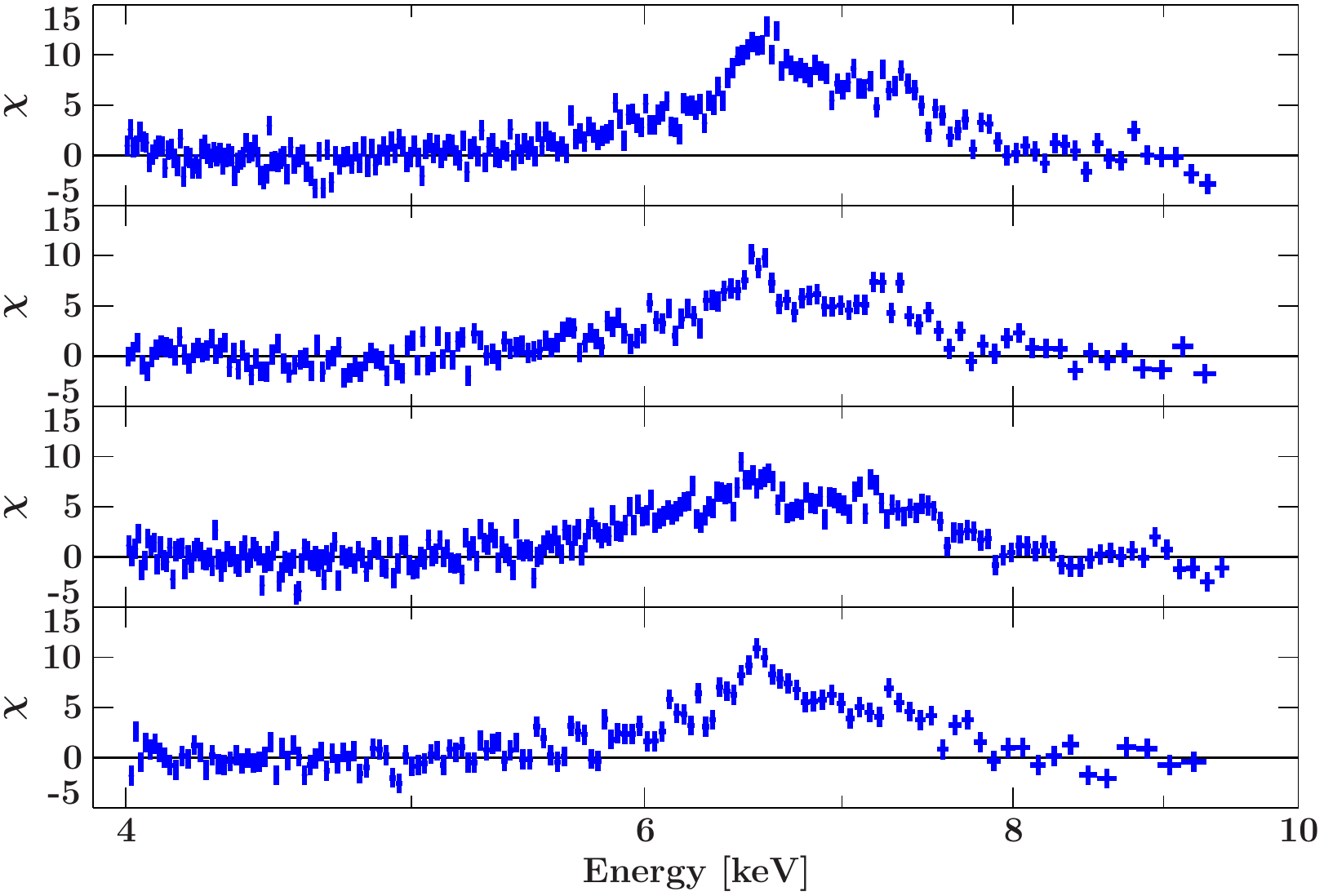}}
  \caption{Broad residuals at the iron line region (around 6.4\,keV)
    in \xmm\,\pn observations are uncovered by excluding the
    5.5--8.0\,keV region and modeling the remaining data with a cutoff
    power-law. There are evident absorption features in the line
    spectra that are due to \ion{Fe}{xxv} (6.65\,keV) and
    \ion{Fe}{xxvi} (6.96\,keV) in the local medium.}
  \label{fig:iron_residuals}
\end{figure}

The \pn data were reduced following standard Timing Mode data
reduction steps with the Science Analysis Software (SAS)
version~11.0.0 and the calibration of the Modified Timing Mode
discussed in Appendix~\ref{app:A}. The increased soft X-ray emission
of the transitional state of the source (Sect.~\ref{sec:source_state})
can potentially lead to pile-up in the center of the point spread
function. Therefore we ignore the innermost three columns. As the new
lower threshold energy limit of 2.8\,keV might introduce possible
transient effects (see Appendix~\ref{app:A}), we restrict the spectral
analysis to energies above 4\,keV. As seen in
Fig.~\ref{fig:iron_residuals}, the signal to noise ratio of the
remaining data is exceptionally good. Throughout this paper we rebin
all \pn data to $\mathrm{S}/\mathrm{N}=100$, which resulted in a
resolution of 40\,eV at 6.4\,keV, oversampling the detector response
by a factor of $\sim4$. All spectral fitting was performed with the
Interactive Spectral Interpretation System
\citep[ISIS;][]{houck:2000,houck:2002,noble:2008}.

\begin{figure*}
  \centering \resizebox{\hsize}{!}{\includegraphics{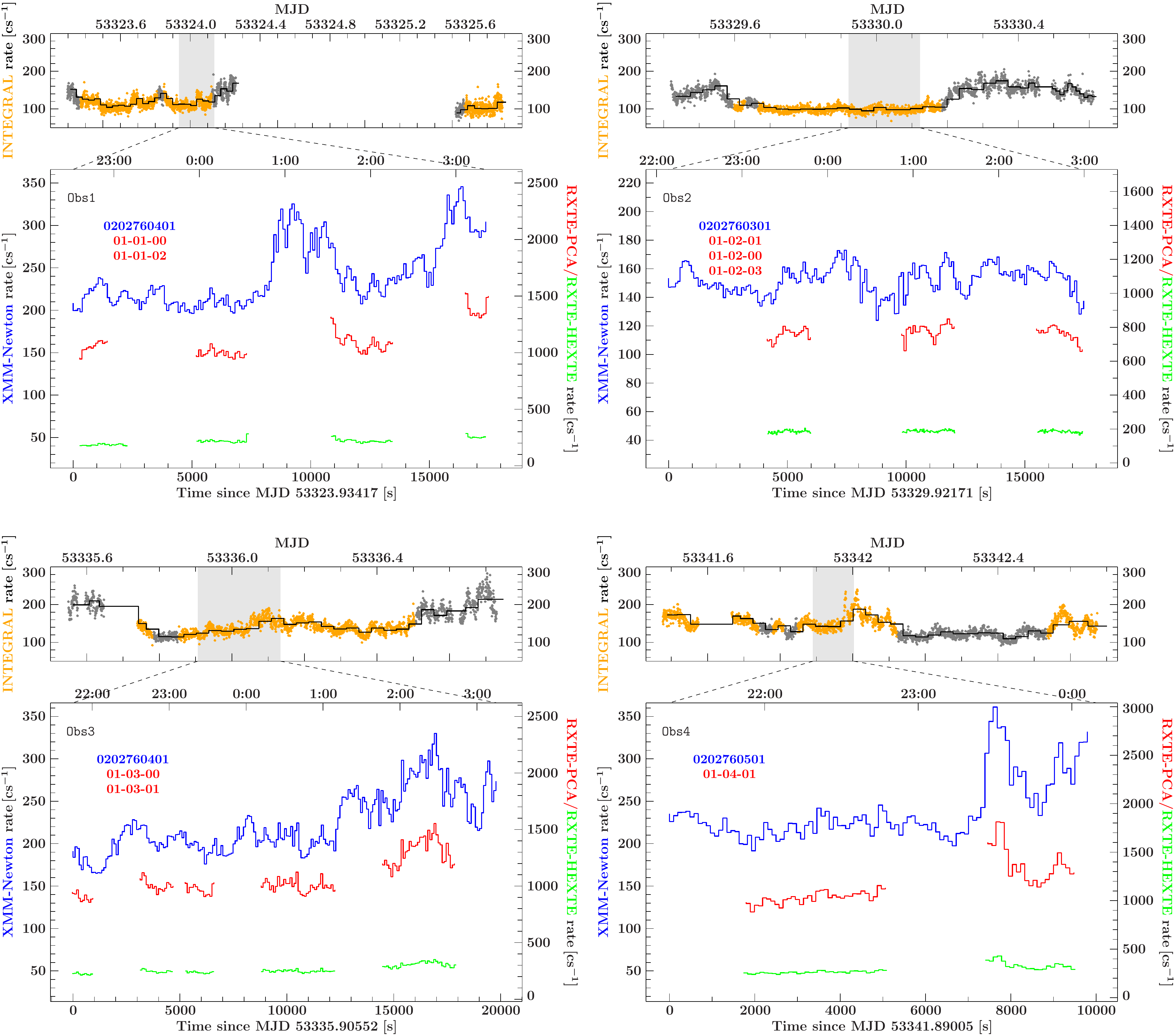}}
  \caption{Background subtracted lightcurves for all four
    observations. Bottom panels: \xmm and \rxte lightcurves,
    binned to a resolution of 96\,s. Top panels:
    \integral-\isgri lightcurves, binned to a resolution of 40\,s. The
    black continuous line shows science window averaged count rates,
    i.e., averages over the $\sim$30\,min long individual INTEGRAL
    pointings. Data used in the analysis, i.e., data with roughly the
    same count rate level as that measured during the \xmm
    observations, are shown in orange. Gray data points are from time
    intervals not used. The same flux variation is evident in all four
    instruments. Note that Obs4 is the shortest observation.
    See Table~\ref{tab:exposure}.}
  \label{fig:xmm_rxte_lc}
\end{figure*}

A preliminary analysis of all four observations, which focused on the
broad band spectrum and the Fe K$\alpha$ line, was published in the
thesis of Sonja Kreykenbohm n\'ee Fritz \citep{fritz:2008}. Compared
to that analysis, the results presented here utilize a significantly
improved calibration strategy and more physical relativistic
reflection models which have been developed since. We started this
analysis with the publication of the observation with the least flux
variability, Obs2 \citep{duro:2011}. Here, we focus on all four
observations, and note the presence of higher flux variability in
observations Obs1, Obs3, and Obs4 (see Fig.~\ref{fig:xmm_rxte_lc} and
Table~\ref{tab:exposure}).

\subsection{\rxte data}\label{sec:rxte}

The broad emission feature (Fig.~\ref{fig:iron_residuals}) makes it
difficult to constrain the underlying continuum using the 4--10\,keV
\xmm modified timing data alone. Therefore we use the much broader
energy-range RXTE data that were taken simultaneously with all four
\xmm observations (observation IDs 90104-01-01-[00,02],
90104-01-02-[00,01,03], 90104-01-03-[00,01] and 90104-04-1;
Fig.~\ref{fig:xmm_rxte_lc}). Its two well-calibrated instruments
provide the coverage up to relatively high energies: the Proportional
Counter Array \citep[PCA;][]{jahoda:2006}, with its five large-area
Proportional Counter Units (PCU), covers the 4--40\,keV range (as used
in this paper), while the High Energy X-ray Timing Experiment
\citep[HEXTE;][]{rothschild:1998} provides us with 20--120\,keV data.

The \rxte data were reduced with HEASOFT~6.11. From the PCA, we use
the photons from the top anode layer of PCU~2 and filter out all data
taken within 10 minutes of passages through the South Atlantic Anomaly
and during the times of high particle background \citep{fuerst:2009}.
A systematic uncertainty of 0.5\% was added in quadrature to the PCA
data in order to account for uncertainties in the calibration. We
applied the same temporal filtering criteria to the \hexte data and
added the spectra from both clusters together for the spectral
analysis. The lightcurves in Fig.~\ref{fig:xmm_rxte_lc} show the
typical \rxte orbital gaps and the flux in both instruments is
following the trends seen in \pn data. \hexte spectra were re-binned
with $\mathrm{S}/\mathrm{N}=30$, and no binning was performed on the
PCA data.

\subsection{Gainshift correction}\label{sec:gainshift}   
As we discussed earlier \citep{duro:2011}, a single continuum model is
not sufficient to describe the 4--120\,keV data. Strong positive
residuals remain in the broad Fe~$\mathrm{K}\alpha$ line region and a
mismatch in the \rxte and \xmm residuals is apparent: the \pn line
appear to be shifted to a slightly higher energy \citep[Fig.~2 of
][]{duro:2011}. This is also evident when adding a neutral 6.4\,keV
iron line to the data, which is required to be at 6.6\,keV by the \pn
data. This value is inconsistent with the \chandra measurements for
the narrow feature in \cyg, which clearly identify a spectral emission
feature at 6.4\,keV \citep[e.g.,][]{miller:2002}.

As discussed in more detail in Appendix~\ref{app:C}, this apparent
difference between \xmm and RXTE is due to Charge Transfer Efficiency
(CTE) effects in the \pn instrument, a well-known effect that occurs
in all X-ray CCDs: As the charge deposited at the impact point of a
X-ray photon is moved to the border of the CCD and read out, some of
the charge is lost to energy sinks (``traps'') in the silicon crystal.
For low count rates the amount of charge lost is well-known as a
function of the impact position of the photon. The standard analysis
pipelines take the known value of the CTE into account when converting
the measured charge back into a ``pulse invariant'' photon energy
\citep[see, e.g.,][]{guainazzi:14,pintore:2014}. When a bright source
is observed, the photon rate can become high enough that the traps
become filled in electrons. As a result, less charge is lost during
the read out of the CCD than in normal observations and the charge
arriving at the read out of the EPIC-pn CCD is larger than during
normal operations. When converting this charge back to a photon energy
in the SAS, the energy assigned to the event, $E_\mathrm{obs}$, will
be wrong. As discussed further in Appendix~\ref{app:C}, this
erroneous energy assignment can be modeled as a linear gain shift,
which can be parameterized with a model of the form
\begin{equation}\label{eq:gainshift}
E_\mathrm{real}=E_\mathrm{obs}/s+\Delta E
\end{equation}
where $E_\mathrm{obs}$ is the energy assigned to events in the SAS,
$E_\mathrm{real}$ is the real photon energy, $s$ is the gain shift
parameter and $\Delta E$ is a potential energy offset, which should be
0\,keV for a purely CTE related gain shift. As we show in
Appendix~\ref{app:C}, joint fitting of narrow line features from
\ion{Fe}{xxv} and \ion{Fe}{xxvi} as seen in the \chandra and \xmm
data, as well as joint fitting of the spectral continuum as measured
with the RXTE-\pca and \pn data, gives consistent values of the energy
gain-shift $s\sim1.02$ to a precision of $\sim2\%$, while the
intercept $\Delta{E}$ is consistently at 0\,keV. Applying this gain shift
correction to the XMM data then achieves a consistent description of
the \xmm-pn and \pca data.

\subsection{\integral data}\label{sec:integral}

In addition to the RXTE observations, we also obtained simultaneous
INTEGRAL data \citep[\integral;][]{winkler:2003} during the \xmm
observations (see Table~\ref{tab:exposure}). We use data from the
INTEGRAL Soft Gamma-Ray Imager \citep[ISGRI;][]{Lebrun:2003} of the
Imager on-Board the INTEGRAL Satellite (IBIS) coded mask instrument
\citet{ubertini:2003} in the 20--500\,keV range.

We reduced the data with OSA~9.2 following the standard reduction
procedures \citep{Goldwurm:2003}, and only used data where the source
was within $14^\circ$ of the center of the field of view of the IBIS
detector. IBIS spectra were extracted in 64 energy channels. As the
strictly simultaneous data have too short exposure times
($t_\mathrm{exp}\sim12$\,ks) to achieve a reasonable quality for
\integral data, we investigated different possibilities for co-adding
of spectra.

Using data from a whole \integral revolution increases the exposure
time to $\sim$70\,ks. However, the joint modeling with \xmm and \rxte
does not give a good description of the data. This is likely due to
the variability of the source during the individual observations
(Fig.~\ref{fig:xmm_rxte_lc}), which are accompanied by changes in
spectral shape during these longer \integral exposures. Note that we
have found that \cyg can undergo spectral changes on time scales as
short as about half an hour \citep{boeck:2011}. The interpretation is
supported by the fact that the use of two power-laws, one for \pn,
\pca and \hexte, and one for the full \isgri exposure, results in
slightly different photon indices, indicating that the spectra have a
slightly different continuum shape.
  
To reduce the influence of this intrinsic source variability and at
the same time to maximize the exposure time, we use spectral data from
those periods within each \integral revolution, where the \isgri flux
levels are comparable to the \isgri flux levels during the strictly
simultaneous \xmm observations (Fig.~\ref{fig:xmm_rxte_lc}). This
approach produces a good overall description of the spectrum.

As a cross check, in order to better assess the influence of \integral
data we modeled the data of all instruments using all three approaches
to \integral data. As expected given the \isgri energy range, the
parameters of the Fe K$\alpha$ line, which is the focus of this paper,
were not significantly influenced by small changes in the high energy
continuum.

\section{The Relativistically Broadened Fe K$\alpha$ line}
\label{sec:relline} 
\subsection{Continuum Model}\label{subsec:coronal_geometry}

The raw data of \pn show the presence of a spectral emission feature
at $\sim$6.4\,keV. To assess the feature we first model the spectrum
excluding bins in the energy range from 5.5--8\,keV with a single
power-law, which is then applied to the complete 4--9.5\,keV region.
Figure~\ref{fig:iron_residuals} shows clear residuals in all four
observations. As discussed earlier \citep{wilms:2006,duro:2011}, the
broad feature is a direct signature of two major effects: the Compton
broadening in the strongly ionized reflector that smears out all other
discrete features such as edges in the reflection spectrum and the
Doppler and general relativistic effects that affect the photons of
the intrinsically narrow Fe K$\alpha$ line and that arise due to the
proximity of the reflector to the black hole. Additionally, our
observations were made during the orbital phase
$\phi_\mathrm{orb}\sim0.8$ so that a fraction of the photons is
absorbed by the local medium \citep{hanke:2009,miskovicova:2013}. This
produces the two absorption lines of \ion{Fe}{xxv} and \ion{Fe}{xxvi}
which are superimposed on the broad emission line (see also
\citealt{diaz:2006} and references therein for a discussion of ionized
absorbers). These lines need to be taken into account during the
spectral modeling. We also generally assume an optically thick and
geometrically thin accretion disk \citep{shakura:1973,novikov:1973},
i.e., a radius-dependent line emissivity per disk unit area that
scales with $r^{-\epsilon}$.

Because of the complex reflection spectrum, modeling the spectrum by
just ading a relativistic line to a continuum described by a power-law
with an exponential cutoff does not result in a good description of
the data. The reduced $\chi^2$ values of such fits vary from 1.44 to
2.17 if $\epsilon$ is left free, and are 4.61 and higher for fixed
$\epsilon=3$. We therefore describe the data with a more physics based
model, which properly takes into account reflection from an ionized
disk. In order to do so, we use the \texttt{reflionx} model
\citep[e.g.,][]{ross:2005,fabian:2010}, which effectively models the
main reflection features: the strongest line transitions, and the
Compton reflection component between 20 and 40\,keV, which is due to
the Compton scattering in the disk. The model's most important
parameters are the Fe abundance and the ionization parameter, $\xi$,
expressed in units of $\mathrm{erg}\,\mathrm{cm}\,\mathrm{s}^{-1}$.

\texttt{Reflionx} does not include the relativistic effects due to the
emitting region's proximity to the black hole. To take them into
account, we convolve the reflection spectrum with the relativistic
convolution model \texttt{relconv} \citep{dauser:2010}. The underlying
spectral continuum is described by a cutoff power-law (note that the
value of 300\,keV that is assumed by \texttt{reflionx} is too high for
the values inferred for \cyg, which yield
$E_\mathrm{fold}\le200\,\mathrm{keV}$). We assume isotropic angular
distribution of the fluorescence photons, which are due to the
up-scattered photons coming from a hot corona \citep{svoboda:2009}. In
addition, the soft excess is described with the multi temperature disk
black body \texttt{diskbb} \citep{mitsuda:1984}, while the absorption
lines of \ion{Fe}{xxv} and \ion{Fe}{xxvi} are described with gaussian
profiles, \texttt{gabs}. The narrow 6.4\,keV emission line feature is
described with a Gaussian model (\texttt{egauss} in ISIS). In all
spectral models flux normalization constants, \texttt{constant} in
ISIS, are given relative to \pca, differences in flux normalization
between the instruments are modeled with instrument-dependent
multiplicative constants. In summary, in ISIS notation the spectral
model is
\begin{multline}
  N_\mathrm{ph}(E)=\mathrm{constant} \cdot
  (\mathrm{gabs}_{1}+\mathrm{gabs}_{2}) \cdot \\
  (\mathrm{cutoffpl}+\mathrm{diskbb}+\mathrm{egauss}+\mathrm{relconv}\otimes\mathrm{reflionx})
\end{multline}
 
\subsection{Continuum}\label{subsec:emm3}
\begin{figure}
  \centering
  \resizebox{\hsize}{!}{\includegraphics{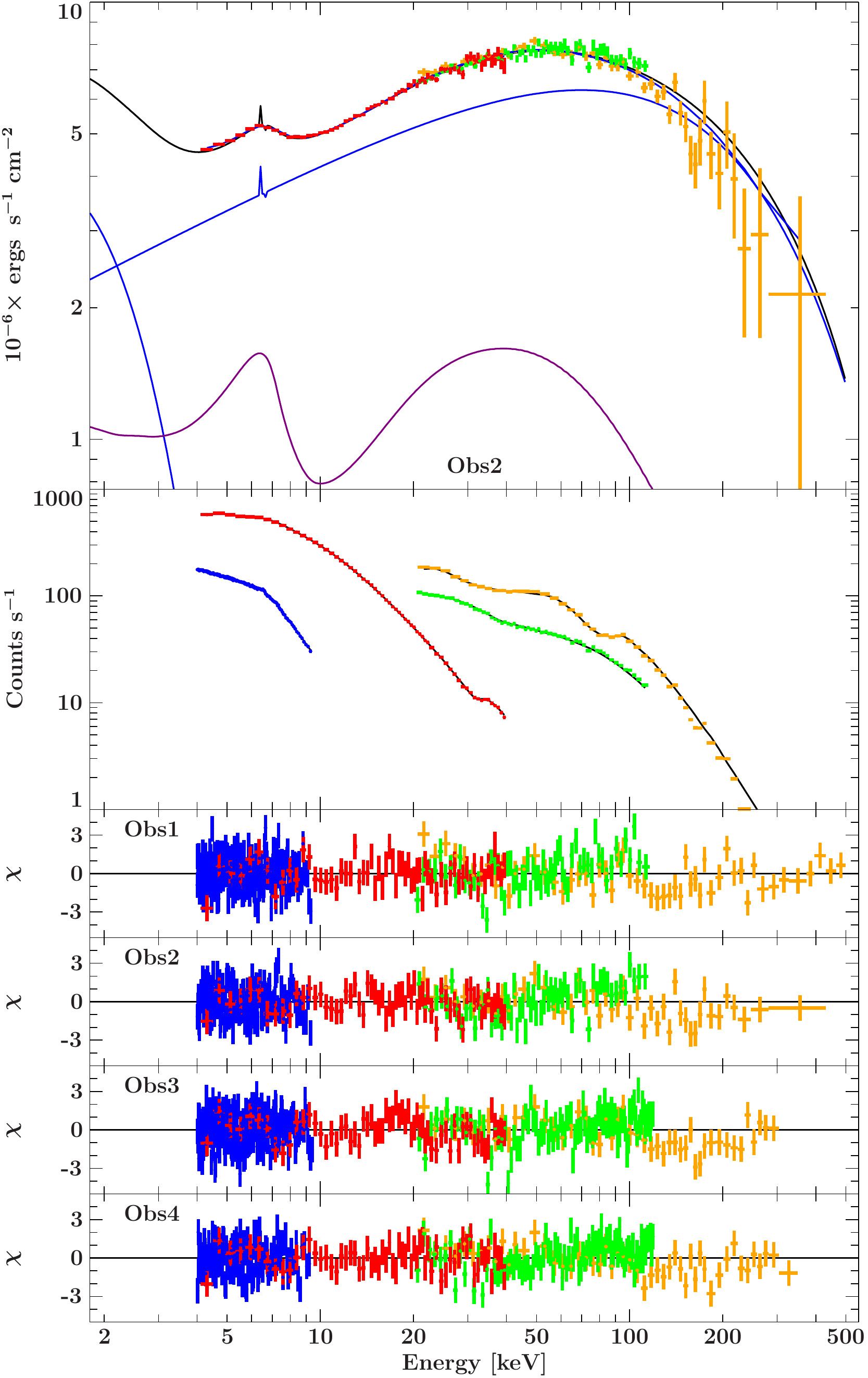}}
  \caption{Spectral modeling of the joint data. Top: Unfolded \xmm,
    \rxte and \integral data and best fit model components for the
    Corona geometry for Obs2. Blue line: continuum spectrum including the
    narrow iron K$\alpha$ emission line and the black body. Purple
    line: relativistically smeared reflection component. Second panel:
    Measured count rate spectra. The four lowest panels show the best
    fit residuals for observations Obs1--Obs4. \pn is blue, \pca is
    red, \hexte is green and \isgri is orange.}
  \label{fig:components}
\end{figure}

Figure~\ref{fig:components} shows that for all four observations the
model describes the 4--500\,keV energy spectrum well. The reduced
$\chi^{2}$ varies from $\chi^2_\mathrm{red}=1.18$ (Obs4; 331\,dof) to
$\chi^2_\mathrm{red}=1.46$ (Obs1, 393\,dof), see
Table~\ref{tab:powEmm_3}. The normalization constants between the
instruments are stable and well constrained.

The photon index shows little change between the observations,
$\Gamma\sim1.6$, with only Obs1 softening to $\Gamma\sim1.7$. In this
observation the source is brighter and the flux strongly variable for
a significant part of this observation\footnote{An increase of $\sim$
  40\%; compare to the stable Obs2, the less variable Obs3, and Obs4
  that only has a short flare.} (Fig.~\ref{fig:xmm_rxte_lc}), probably
leading to the softening and the comparatively poor fits. This
interpretation is consistent with the highest gain-shift value for
Obs1 ($s_\mathrm{gainshift}\sim1.027$) as the high source flux is able
to saturate silicon traps faster, leading to a wrong CTE correction
(see Appendix~\ref{app:C}). Excluding the period of increased flux and
fitting Obs1 again, using a total of 8\,ks of \pn observation with the
corresponding \hexte, \pca and same flux \isgri data
(Fig.~\ref{fig:xmm_rxte_lc}), results in an excellent spectral model
with $\chi^{2}_\mathrm{red}\sim1.13$. The gain-shift slope decreases
in value to $s_\mathrm{gainshift}=1.0187^{+0.0020}_{-0.0017}$, in
better agreement with other observations, i.e., the CTE
over-correction is less severe. The photon index hardens to
$\Gamma=1.694^{+0.020}_{-0.016}$ confirmining the effect of the flare
on the overall shape of the continuum that leads to the bad fit to the
total Obs1.

\begin{figure}
  \centering
  \resizebox{\hsize}{!}{\includegraphics{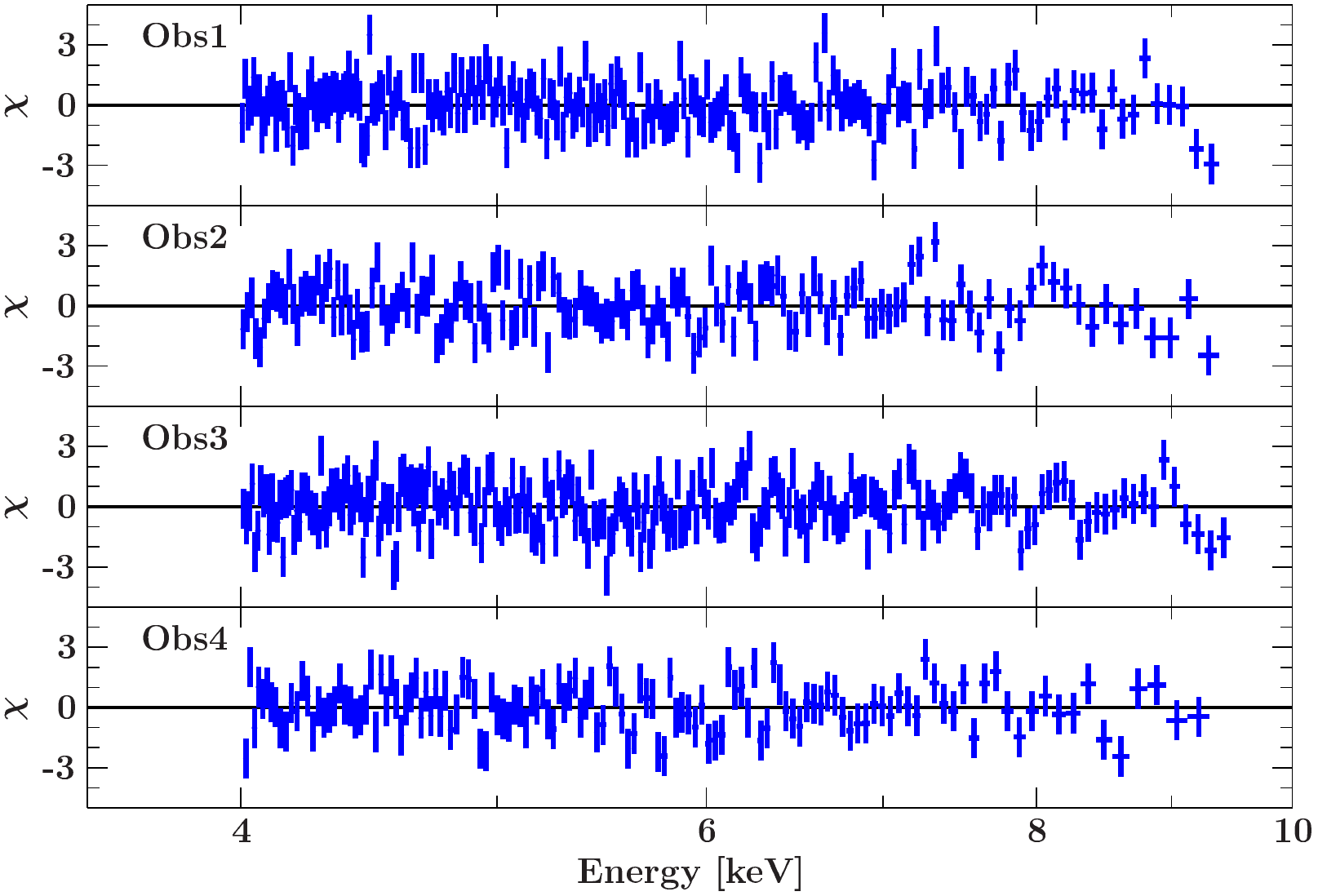}}
  \caption{Best fit \xmm residuals for Corona geometry for observations
    Obs1--Obs4 for the fits of Fig.~\ref{fig:components}. No
    systematic variations are present.}
  \label{fig:xmm_systematics}
\end{figure}

The folding energy, which reflects the electron temperature in the
corona, is stable for all four observations around
$E_\mathrm{fold}\,\sim\,170\,\mathrm{keV}$. The study by
\citet{nowak:2011} that uses hard-state broad band data of \suzaku and
\rxte shows folding energy values that are consistent with our results
\citep[see also][]{wilms:2006b}.

A possible criticism of our model is that it does not fit well the
data above 100\,keV, where \isgri is dominant
(Fig.~\ref{fig:components}). As already discussed in
Sect.~\ref{sec:integral}, using only strictly simultaneous \isgri data
does not improve the results. We expect the broad band \isgri data to
mostly affect hard continuum parameters, especially the folding energy
$E_\mathrm{fold}$. We can observe exactly this behavior in our data:
repeating the modeling excluding the \isgri data, the value
$E_\mathrm{fold}$ strongly increases and it is weakly constrained
\citep[$E_\mathrm{fold}=290^{+80}_{-50}$ keV,][]{duro:2011}, while the
remaining parameters do not change significantly.

Furthermore, it is clear that the underlying continuum affects the fit
to the Fe K$\alpha$ line in general. It is therefore important to see
that the fit to the continuum does not produce any systematic
variations in the residuals of the \xmm data, which are critical for
the extraction of line parameters. As Fig.~\ref{fig:xmm_systematics}
shows, no significant systematic variations can be observed.

\subsection{Coronal geometry}
\subsubsection{Disk reflection with $\epsilon=3$}\label{sec:refl3}

Having established that our continuum parameters are largely stable
throughout all four observations, we now focus on the disk reflection.
We start with modeling the disk reflection using the commonly used
empirical emissivity profile $\propto r^{-\epsilon}$ with a standard
value of $\epsilon=3$ \citep[see, e.g.,][]{fabian:2012}.
Figure~\ref{fig:components} and Model~1 in Table~\ref{tab:powEmm_3}
show the resulting best fit residuals and values. The line profile
indicates a high angular momentum of the black hole. All four
observations point consistently towards a spin value of $a>0.9$. In
Fig.~\ref{fig:contPL} we examine the correlation between the
parameters with the largest impact on the shape of the Fe K$\alpha$
line feature, i.e., spin, $a$, inclination, $i$, and emissivity index,
$\epsilon$. The third row of Fig.~\ref{fig:contPL} shows correlations
with models where $\epsilon$ was held fixed at $\epsilon=3$. They
result in a high spin solution and an inclination that is consistent
with measurements of $\sim27^\circ$ \citep{orosz:2011}\footnote{If
  only the non-flaring part of Obs1 is considered, the best fit is
  also consistent with this value of the inclination.}. These inferred
spin values agree well with \citet{gou:2011}, who obtain an extreme
spin value for \cyg, $a\sim0.9$, by thermal continuum spectrum
fitting, and with \citet{fabian:2012}, who fit the reflection spectra.
The Fe abundance is well constrained in all cases to a few times the
solar value, consistent with results obtained previously.

\begin{figure}
  \centering
  \resizebox{\hsize}{!}{\includegraphics{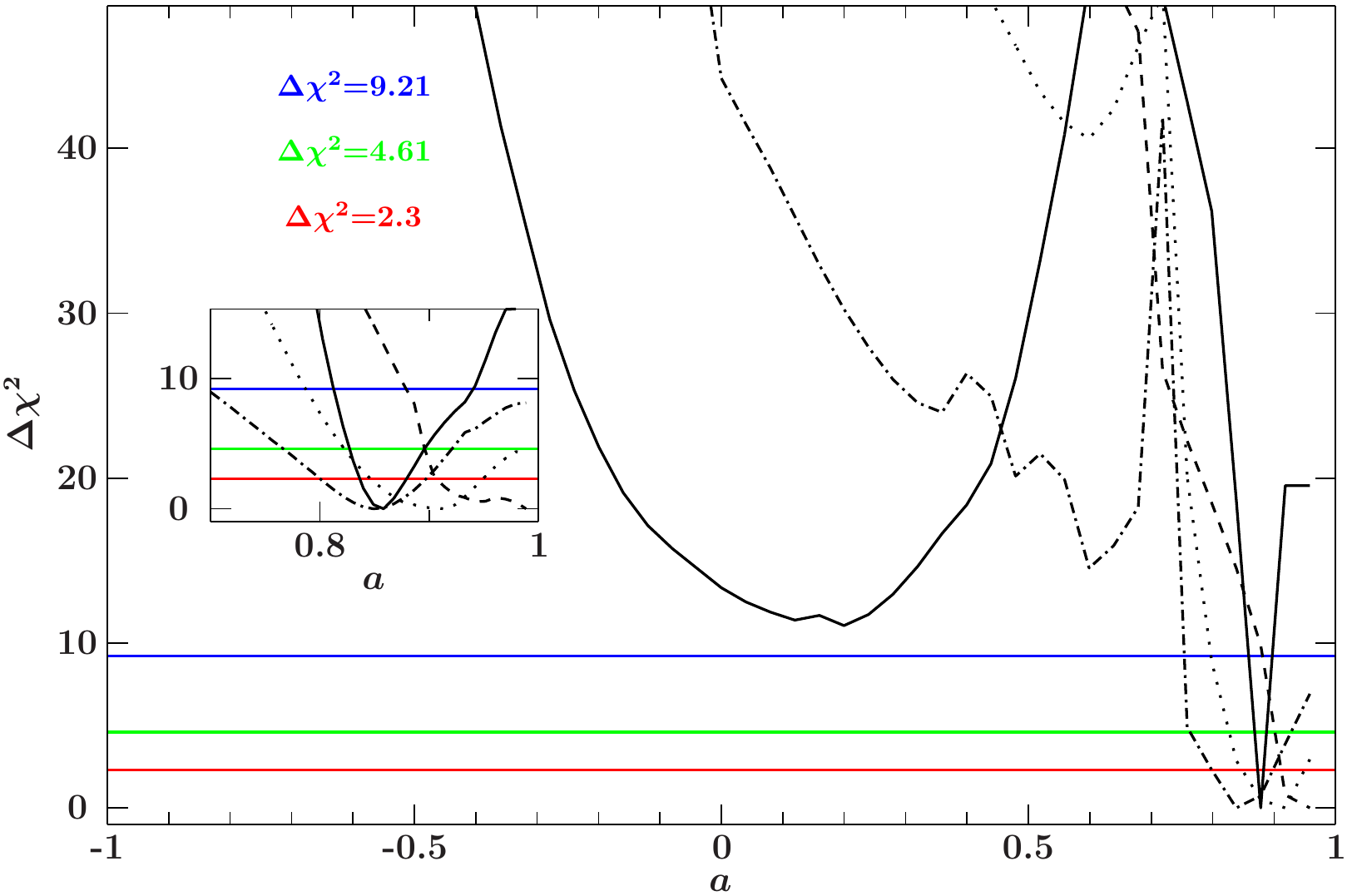}}
  \caption{$\chi^{2}$ behavior of the spin $a$ for the broken power
    law emissivity disk profile (Model~3). All four observations
    have their best minima at $a\sim0.9$. Obs1 is drawn with solid,
    Obs2 with dashed, Obs3 with dotted and Obs4 dot-dash-dotted line.
    The zoom-in figure shows the area around the minima, produced with
    very fine binning. Similar behavior is produced for the single
    power-law profile.}
  \label{fig:steppar}
\end{figure}

\begin{figure*}
  \centering
  \resizebox{17cm}{!}{\includegraphics{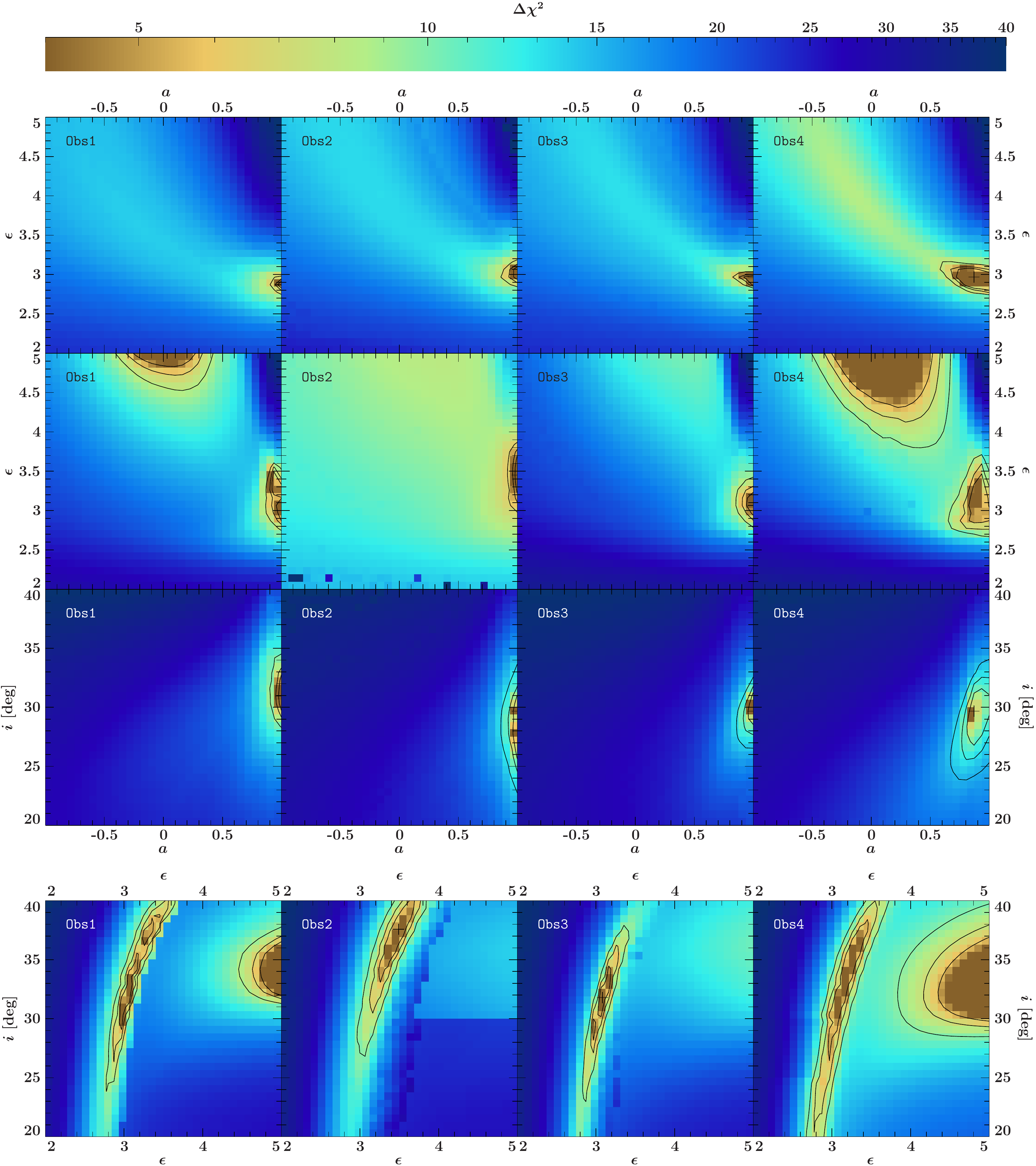}}
  \caption{Parameter correlations for the coronal model with a cutoff
    power-law and relativistic reflection with $\epsilon$ fixed at
    $\epsilon=3$ and as a free parameter (Models~1 and~2 in
    Table~\ref{tab:powEmm_3}). First row: Holding the inclination
    fixed at $i=27^\circ$ results in one best fit solution for the
    emissivity index $\epsilon$ and spin $a$. Second row: for $i$ let
    free, one more solution emerges. We chose the physical one as the
    best fit (see Sect.~\ref{subsec:emmFre}). Third row: $\epsilon=3$
    is chosen to describe the correlation between $i$ and $a$. Fourth
    row: interplay between $\epsilon$ and $i$. Shown are the
    $\chi^{2}$ significance contours for two parameters of interest
    based on $\Delta\chi^{2}=2.30$, 4.61 and 9.21, i.e., 68\%, 90\%
    and 99\% confidence levels, with respect to the best fit case.
  }\label{fig:contPL}
\end{figure*}

\subsubsection{Disk reflection with $\epsilon$ free}\label{subsec:emmFre}

What happens if we leave both the emissivity profile index $\epsilon$
and the inclination $i$ free? Such a setup produces a slightly better
reduced $\chi^{2}$ then a fixed emissivity, ranging from 1.18 (331
dof) to 1.38 (392 dof; Model~2 in Table~\ref{tab:powEmm_3}). In Obs2
and Obs3 the only solution found is for $\epsilon\sim3$ and spin
$a\sim0.9$, reproducing the result for fixed value of $\epsilon$. This
is not entirely true for the observations Obs1 and Obs4 in which the
best fit solution produces retrograde spin values, $a\leq0$. A
negative spin means that the \texttt{ISCO} is further away from the
black hole, with $R_\mathrm{ISCO}\ge 6 r_\mathrm{g}$. Coupled with the
very dramatic increase in the emissivity profile index value
$\epsilon\sim10$ (see Table~\ref{tab:powEmm_3}), this results in an
unphysical situation that is, at best, very difficult to explain. We
note, however, that even for these two observations, we find
acceptable fits for the combination of high $a$ and low $\epsilon$
(see the second row on Fig.~\ref{fig:contPL} for $\chi^2$ significance
maps).

The shape of the line profile is in part determined by $\epsilon$ and
$i$. The higher emissivity index $\epsilon$ for spinning black holes
produces broader and more red-shifted profiles, as the irradiation is
more concentrated on the inner parts of the accretion disk, while the
decreasing inclination $i$ shifts the blue wing in the reflection
feature towards lower energies \citep[see, e.g.,][]{dauser:2013}.
Thus, the emissivity profile index $\epsilon$ and the inclination $i$
are somewhat correlated, as can be seen in the fourth row of
Fig.~\ref{fig:contPL}. On the same figure we see that the inclination
obtained in our fits is consistent with $i\sim27^\circ$. Fixing $i$ at
this value and re-fitting enables us to almost reproduce the results
obtained for $\epsilon$ fixed at a value of~3. The best fit values for
Model~2 are shown in Table~\ref{tab:powEmm_3}.

\subsubsection{Disk Reflection with a broken power-law for the
  emissivity profile}\label{subsec:broken}

In order to improve the unrealistic combinations of low spin and high
emissivity index obtained for a single power-law emissivity
(Sect.~\ref{subsec:emmFre}), we employ a broken power-law emissivity
profile to model the irradiation of the disk in a more realistic way.
For example, a compact corona or an irradiation of the disk by a
compact central source lying above the accretor produces a steep
emissivity profile at the inner part of the disk, converging to
$r^{-3}$ at the outer regions
\citep{fukumura:2007,wilkins:2011,dauser:2013}. To describe this kind
of behavior we describe the emissivity profile with a broken power
law, where we fit the inner emissivity index, $\epsilon_\mathrm{in}$,
and fix the emissivity for the outer disk to
$\epsilon_\mathrm{out}=3$. We constrain the break of the power-law to
occur within $6r_\mathrm{g}$ from the black hole, as no steep
emissivities are expected at larger radii
\citep[e.g.,][]{dauser:2013}. Model~3 in Table~\ref{tab:powEmm_3}
shows the best fit values for all four observations. The table reveals
that $\chi^{2}_\mathrm{red}$ is similar or better than for the single
power-law emissivity profile disk description (with one less degree of
freedom). The best fit residuals are similar to the ones presented in
Fig.~\ref{fig:components} and are therefore not shown.

\begin{figure*}
  \centering
  \resizebox{17cm}{!}{\includegraphics{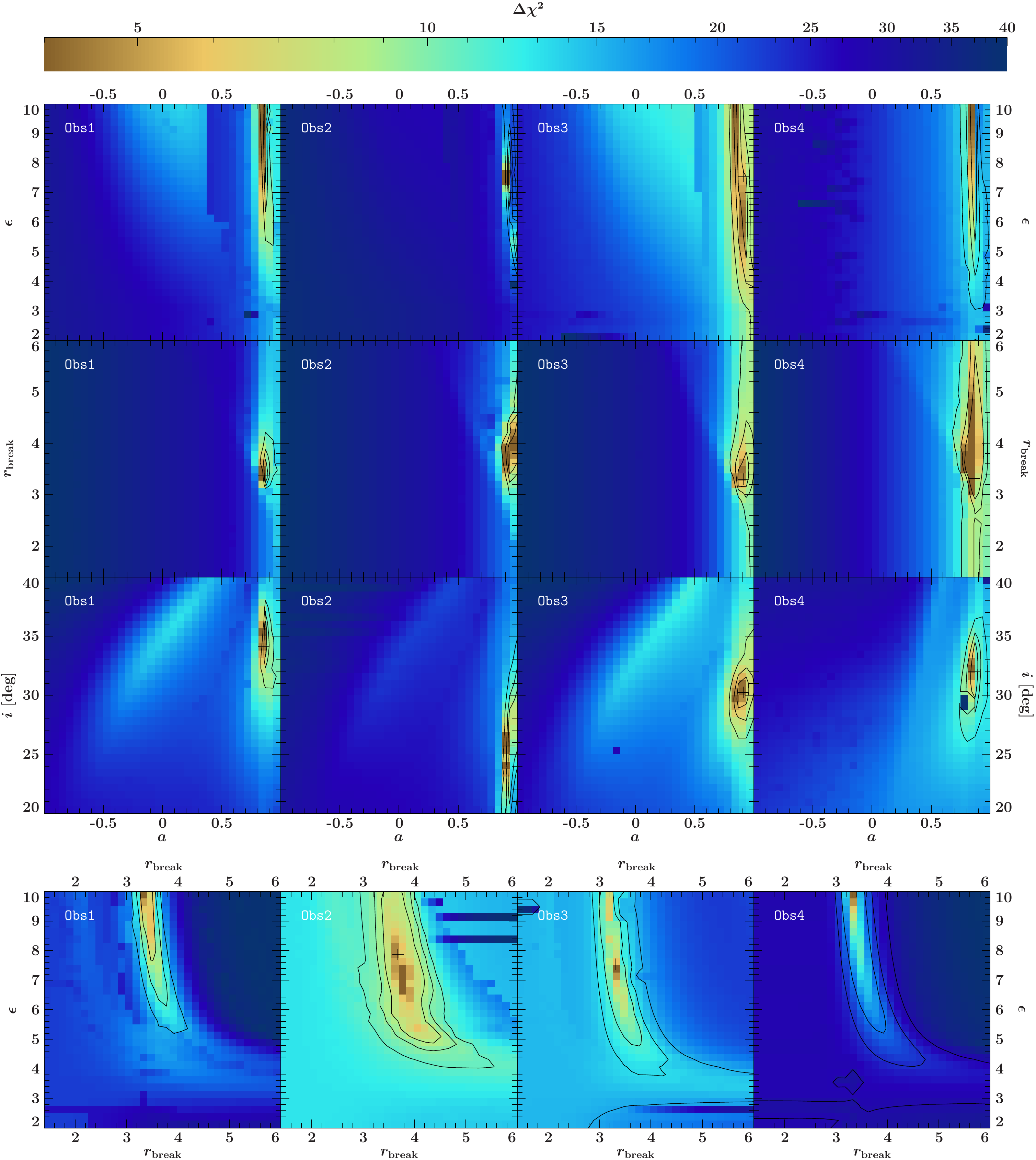}}
  \caption{Parameter correlations for fits with a broken power-law
    emissivity (Model~3 in Table~\ref{tab:powEmm_3}). High spin is
    required in all situations. First row: Emissivity index $\epsilon$
    and spin $a$. Second row: Break radius $r_\mathrm{break}$ versus
    the spin $a$. Third row: Inclination $i$ with respect to the spin
    $a$. Fourth row: Change in emissivity profile index $\epsilon$
    with varying break radius $r_\mathrm{break}$. Shown are the
    $\chi^{2}$ significance contours for two parameters of interest
    based on $\Delta\chi^{2}=2.30$, 4.61, and 9.21, i.e., 68\%, 90\%,
    and 99\% confidence, with respect to the best fit case. }
  \label{fig:contBPL}
\end{figure*}

Compared to the single power-law emissivity description
(Sect.~\ref{subsec:emmFre}), the most prominent change is the
disappearance of the unphysical solutions for Obs1 and Obs4. For all
four observations, only one spin solution with $a\sim0.9$ is found, as
clearly seen on Fig.~\ref{fig:steppar} where we show the behavior of
the goodness of the fit for different values of $a$. The zoom-in on
the high spin region shows that the minima of $\chi^2$ are
consistently at $a\sim0.9$. Note that a similar behavior is achieved
for the single power-law emissivity if we fix $\epsilon=3$. As already
motivated above, the break in emissivity occurs close to the black
hole as expected from calculations of more physical emissivity
profiles \citep[see, e.g.,][]{dauser:2013}. Moreover, the inferred
value of $r_\mathrm{break}\sim3.5r_\mathrm{g}$ is consistent with the
break radius found by \citet{fabian:2012} for \cyg. We find a very
steep emissivity profile for small radii,
$4\le\epsilon_\mathrm{in}\le10$, implying that most of the reflected
Fe K$\alpha$ emission of \cyg comes from these inner regions. The
outer parts are following the standard emissivity profile of a thin
disk. This is evident in the fourth row of Fig.~\ref{fig:contBPL}
where $\epsilon$ changes follow this description with the increasing
distance from the black hole, expressed by the break radius
$r_\mathrm{break}$.

We emphasize that Fig.~\ref{fig:contBPL} clearly shows that the high
spin solution $a\sim0.9$ is required as a best fit solution with
respect to $\epsilon$, $r_\mathrm{break}$, and $i$. This result is
consistent for all four observations. The remaining parameter values
are in agreement with the single power-law emissivity description.
   
\begin{table*}
  \caption{Best fit parameters for a coronal model using a cutoff
    power-law and relativistic reflection. Three situations are
    described (from the top to the bottom): 1) a single power-law
    emissivity profile with fixed emissivity index $\epsilon=3$, 2)
    power-law emissivity with $\epsilon$ free and free inclination
    $i$, 3) a broken power-law emissivity profile with
    $\epsilon_\mathrm{in}$ free and $\epsilon_\mathrm{out}=3$.
    Uncertainties are at the 90\% confidence level for one parameter
    of interest.}\label{tab:powEmm_3}

  \centering
\renewcommand{\arraystretch}{1.19}
\begin{tabular}{ll|llll} 
    \hline 
    \multirow{3}{*}{Model~1}& & Obs1  & Obs2 & Obs3  & Obs4  \\ 
    \cline{2-6}  
    &   $A_\mathrm{pl}$   &   $1.58^{+0.12}_{-0.10}$   &   $1.10^{+0.06}_{-0.09}$   &   $1.53^{+0.08}_{-0.09}$   &   $1.56\pm0.10$    \\ 
    &   $\Gamma_\mathrm{pl}$   &   $1.724\pm0.012$   &   $1.611^{+0.013}_{-0.016}$   &   $1.609^{+0.012}_{-0.010}$   &   $1.577^{+0.015}_{-0.012}$    \\ 
    & $E_\mathrm{{fold}}$  [keV]   &   $181\pm14$   &   $175\pm13$   &   $181^{+12}_{-8}$   &   $164^{+9}_{-8}$    \\ 
    \cline{2-6}  
    &   $A_\mathrm{disk}$   &   $\left(161^{+8}_{-7}\right)\times10^{2}$   &   $\left(69\pm7\right)\times10^{2}$   &   $\left(117^{+7}_{-8}\right)\times10^{2}$   &   $\left(147^{+12}_{-13}\right)\times10^{2}$    \\ 
    \cline{2-6}  
    &   $A_\mathrm{ref}$  [$10^{-5}$]    &   $1.39^{+0.14}_{-0.16}$   &   $1.02^{+0.18}_{-0.13}$   &   $1.33^{+0.20}_{-0.12}$   &   $1.30^{+0.29}_{-0.17}$    \\ 
    &   $\mathrm{Fe}/\mathrm{Fe}_\odot$   &   $3.5^{+0.7}_{-0.6}$   &   $3.1^{+0.8}_{-0.7}$   &   $3.8^{+0.6}_{-0.7}$   &   $3.6^{+1.0}_{-0.7}$    \\ 
    \multirow{1}{*}{$\epsilon=3$}& $\xi$  $[\mathrm{erg}\,\mathrm{cm}\,\mathrm{s}^{-1}]$   &   $2330^{+200}_{-190}$   &   $2040^{+250}_{-380}$   &   $2080^{+170}_{-280}$   &   $2100^{+250}_{-410}$    \\ 
    \cline{2-6}  
    &   $\epsilon$   &   3.0   &   3.0   &   3.0   &   3.0    \\ 
    &   $a$   &   $0.998^{+0.000}_{-0.029}$   &   $0.0.998^{+0.000}_{-0.060}$   &   $0.998^{+0.000}_{-0.060}$   &   $0.89^{+0.12}_{-0.09}$    \\ 
    &   $i$ [deg]   &   $31.1^{+1.9}_{-1.5}$   &   $27.8^{+3.6}_{-2.4}$   &   $29.8^{+1.7}_{-1.5}$   &   $29.6^{+2.2}_{-3.1}$    \\ 
    \cline{2-6}  
    &   $c_\mathrm{HEXTE}$   &   $0.843\pm0.006$   &   $0.826\pm0.006$   &   $0.834\pm0.004$   &   $0.826\pm0.005$    \\ 
    &   $c_\mathrm{PN}$   &   $0.8377^{+0.0049}_{-0.0029}$   &   $0.780^{+0.008}_{-0.004}$   &   $0.8139^{+0.0026}_{-0.0029}$   &   $0.813\pm0.004$    \\ 
    &   $c_\mathrm{ISGRI}$   &   $0.922^{+0.010}_{-0.009}$   &   $0.921\pm0.009$   &   $0.903\pm0.008$   &   $0.905\pm0.008$    \\ 
    \cline{2-6}  
    &   $s_\mathrm{gainshift}$   &   $1.0266^{+0.0010}_{-0.0023}$   &   $1.0180^{+0.0020}_{-0.0044}$   &   $1.0184^{+0.0012}_{-0.0008}$   &   $1.0209^{+0.0019}_{-0.0012}$    \\ 
    \cline{2-6}  
    &   $\chi^{2}/\mathrm{dof}$ &  574.0/393&  418.7/320&  582.1/443&  406.1/332   \\ 
    &   $\chi^{2}_\mathrm{red}$ &  1.46&  1.31&  1.31&  1.22   \\ 
    \hline 
    \hline
Model~2    & $A_\mathrm{pl}$   &   $1.77^{+0.08}_{-0.12}$   &   $1.20^{+0.07}_{-0.08}$   &   $1.57^{+0.07}_{-0.10}$   &   $1.62^{+0.05}_{-0.10}$    \\ 
    & $\Gamma_\mathrm{pl}$   &   $1.736\pm0.013$   &   $1.630^{+0.017}_{-0.018}$   &   $1.611^{+0.013}_{-0.010}$   &   $1.583^{+0.010}_{-0.013}$    \\ 
    & $E_\mathrm{{fold}}$  [keV]   &   $172^{+12}_{-10}$   &   $181\pm16$   &   $179^{+12}_{-9}$   &   $163^{+9}_{-5}$    \\ 
    \cline{2-6}  
    & $A_\mathrm{disk}$   &   $\left(187^{+10}_{-12}\right)\times10^{2}$   &   $\left(62^{+8}_{-7}\right)\times10^{2}$   &   $\left(115\pm8\right)\times10^{2}$   &   $\left(155^{+14}_{-16}\right)\times10^{2}$    \\ 
    \cline{2-6}
    & $A_\mathrm{ref}$  [$10^{-5}$]    &   $1.33^{+0.19}_{-0.13}$   &   $1.6\pm0.4$   &   $1.36^{+0.35}_{-0.14}$   &   $1.30^{+0.20}_{-0.18}$    \\ 
    & $\mathrm{Fe}/\mathrm{Fe}_\odot$   &   $2.7^{+0.6}_{-0.5}$   &   $3.9\pm0.9$   &   $3.9^{+0.7}_{-0.8}$   &   $3.3^{+0.8}_{-0.6}$    \\ 
    \multirow{1}{*}{$\epsilon=\mathrm{free}$}&$\xi$  $[\mathrm{erg}\,\mathrm{cm}\,\mathrm{s}^{-1}]$   &   $2030^{+180}_{-310}$   &   $1340^{+610}_{-190}$   &   $2010^{+210}_{-410}$   &   $1930^{+280}_{-260}$    \\ 
    \cline{2-6}  
    & $\epsilon$   &   $10.0^{+0.0}_{-1.7}$   &   $3.49^{+0.25}_{-0.31}$   &   $3.08^{+0.19}_{-0.17}$   &   $10^{+0}_{-4}$    \\ 
    & $a$   &   $-0.59\pm0.20$   &   $0.998^{+0.000}_{-0.028}$   &   $0.998^{+0.000}_{-0.080}$   &   $-0.36^{+0.30}_{-0.24}$    \\ 
    & $i$ [deg]   &   $33.4^{+1.0}_{-0.8}$   &   $37.6^{+2.5}_{-5.5}$   &   $32^{+4}_{-5}$   &   $33.1^{+1.5}_{-1.4}$    \\ 
    \cline{2-6}  
    & $c_\mathrm{HEXTE}$   &   $0.845^{+0.006}_{-0.005}$   &   $0.825\pm0.006$   &   $0.834\pm0.004$   &   $0.827\pm0.005$    \\ 
    & $c_\mathrm{PN}$   &   $0.8436^{+0.0057}_{-0.0029}$   &   $0.780\pm0.005$   &   $0.8141^{+0.0026}_{-0.0033}$   &   $0.812\pm0.004$    \\ 
    & $c_\mathrm{ISGRI}$   &   $0.924\pm0.009$   &   $0.919^{+0.010}_{-0.009}$   &   $0.903\pm0.008$   &   $0.906\pm0.008$    \\ 
    \cline{2-6}  
    & $s_\mathrm{gainshift}$   &   $1.0205^{+0.0009}_{-0.0025}$   &   $1.0185^{+0.0027}_{-0.0030}$   &   $1.0184^{+0.0013}_{-0.0008}$   &   $1.0204^{+0.0014}_{-0.0018}$    \\ 
    \cline{2-6}  
    & $\chi^{2}/\mathrm{dof}$ &  541.1/392&  411.0/319&  581.4/442&  391.3/331   \\ 
    & $\chi^{2}_\mathrm{red}$ &  1.38&  1.29&  1.32&  1.18   \\ 
    \hline 
    \hline 
Model~3    & $A_\mathrm{pl}$   &   $1.88\pm0.08$   &   $1.08\pm0.10$   &   $1.60^{+0.05}_{-0.10}$   &   $1.66^{+0.08}_{-0.07}$    \\ 
    & $\Gamma_\mathrm{pl}$   &   $1.748\pm0.014$   &   $1.600^{+0.022}_{-0.019}$   &   $1.614^{+0.009}_{-0.013}$   &   $1.587^{+0.016}_{-0.015}$    \\ 
    & $E_\mathrm{{fold}}$  [keV]   &   $177^{+14}_{-13}$   &   $170\pm9$   &   $180^{+10}_{-9}$   &   $164^{+12}_{-10}$    \\ 
    \cline{2-6}    
    & $A_\mathrm{disk}$   &   $\left(163\pm10\right)\times10^{2}$   &   $\left(47^{+9}_{-27}\right)\times10^{2}$   &   $\left(97^{+20}_{-23}\right)\times10^{2}$   &   $\left(138^{+15}_{-20}\right)\times10^{2}$    \\ 
    \cline{2-6}  
    & $A_\mathrm{ref}$  [$10^{-5}$]    &   $1.83^{+0.30}_{-0.31}$   &   $1.13^{+0.35}_{-0.12}$   &   $1.36^{+0.22}_{-0.17}$   &   $1.6^{+0.5}_{-0.4}$    \\ 
    & $\mathrm{Fe}/\mathrm{Fe}_\odot$   &   $3.6^{+0.7}_{-0.4}$   &   $6.0^{+0.0}_{-1.7}$   &   $4.4^{+1.6}_{-1.0}$   &   $4.3^{+1.3}_{-0.8}$    \\ 
    \multirow{1}{*}{$\epsilon_\mathrm{in}=\mathrm{free}$}& $\xi$  $[\mathrm{erg}\,\mathrm{cm}\,\mathrm{s}^{-1}]$   &   $1570^{+330}_{-230}$   &   $2200^{+400}_{-500}$  &   $2000^{+310}_{-290}$   &   $1700^{+500}_{-400}$    \\ 
    \cline{2-6}  
    \multirow{1}{*}{$\epsilon_\mathrm{out}=3$}& $\epsilon_\mathrm{in}$   &   $10.0^{+0.0}_{-3.0}$   &   $5.4^{+4.6}_{-0.7}$   &   $7.5^{+2.5}_{-3.1}$   &   $10^{+0}_{-6}$    \\ 
    & $\epsilon_\mathrm{out}$   &   3.0   &   3.0   &   3.0   &   3.0    \\ 
    & $r_\mathrm{br}$  $[GM/c^2]$   &   $3.38^{+0.27}_{-0.15}$   &   $4.0^{+0.7}_{-0.6}$   &   $3.3^{+0.7}_{-0.4}$   &   $3.31^{+0.78}_{-0.24}$    \\ 
    & a   &   $0.856^{+0.026}_{-0.020}$   &   $0.989^{+0.009}_{-0.088}$   &   $0.91^{+0.05}_{-0.07}$   &   $0.86\pm0.05$    \\ 
    & $i$  $[\mathrm{deg}]$   &   $34.1^{+2.4}_{-1.8}$   &   $28\pm4$   &   $30.2^{+1.6}_{-2.5}$   &   $32.0^{+2.8}_{-2.9}$    \\ 
    \cline{2-6}    
    & $c_\mathrm{HEXTE}$   &   $0.844\pm0.006$   &   $0.823\pm0.007$   &   $0.834\pm0.004$   &   $0.826\pm0.005$    \\ 
    & $c_\mathrm{PN}$   &   $0.838^{+0.010}_{-0.004}$   &   $0.782^{+0.007}_{-0.004}$   &   $0.8134^{+0.0030}_{-0.0036}$   &   $0.813\pm0.004$    \\ 
    & $c_\mathrm{ISGRI}$   &   $0.923\pm0.009$   &   $0.916\pm0.010$   &   $0.903\pm0.008$   &   $0.905\pm0.008$    \\ 
    \cline{2-6}    
    & $s_\mathrm{gainshift}$   &   $1.0251^{+0.0016}_{-0.0010}$   &   $1.0225^{+0.0023}_{-0.0032}$   &   $1.0184^{+0.0015}_{-0.0007}$   &   $1.0213^{+0.0018}_{-0.0015}$    \\ 
    \cline{2-6}    
    & $\chi^{2}/\mathrm{dof}$ &  544.6/386&  394.6/318&  574.3/441&  397.7/330   \\ 
    & $\chi^{2}_\mathrm{red}$ &  1.41&  1.24&  1.30&  1.21   \\ 
    \hline
\end{tabular}

\end{table*}

\subsection{Lamp post geometry}
\label{subsect:lp}

\begin{figure}
  \centering
  \resizebox{\hsize}{!}{\includegraphics{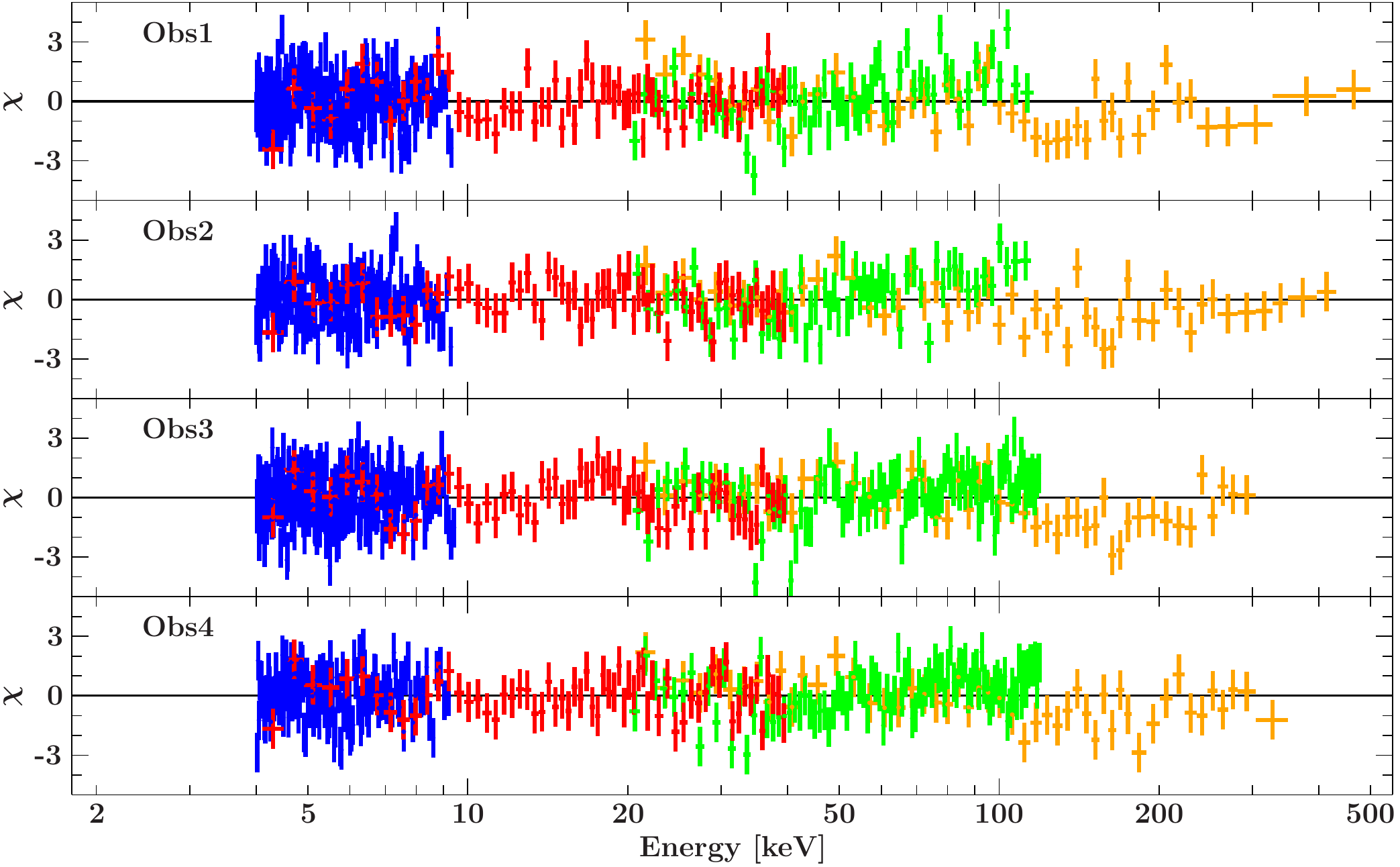}}
  \caption{Residuals for fits with the lamp post model for all four
    observations using \xmm (blue), \rxte (\pca: red, \hexte: green),
    and \integral (orange) data. }
  \label{fig:residuals_lp}
\end{figure}

\begin{table*}
  \caption{Best fit parameters for the lamp post model for all four
    observations (Model~4). Uncertainties are at the 90\% level for
    one interesting parameter.}\label{tab:lp} \centering
\renewcommand{\arraystretch}{1.2}
\begin{tabular}{lllll} 
 \hline 
Parameter  & Obs1  & Obs2  & Obs3  & Obs4  \\ 
\hline  
$A_\mathrm{pl}$   &   $1.91^{+0.10}_{-0.09}$   &   $1.21^{+0.07}_{-0.08}$   &   $1.58^{+0.07}_{-0.06}$   &   $1.67^{+0.10}_{-0.14}$    \\
$\Gamma_\mathrm{pl}$   &   $1.754^{+0.015}_{-0.013}$   &   $1.622^{+0.029}_{-0.017}$   &   $1.614^{+0.014}_{-0.009}$   &   $1.591^{+0.017}_{-0.020}$    \\ 
$E_\mathrm{{fold}}$  [keV]   &   $181^{+15}_{-13}$   &$171^{+29}_{-12}$   &   $180\pm10$   & $166\pm10$  \\ 
\hline  
$A_\mathrm{disk}$   &   $\left(164^{+10}_{-14}\right)\times10^{2}$   &$\left(54^{+10}_{-12}\right)\times10^{2}$   &$\left(116^{+8}_{-9}\right)\times10^{2}$&$\left(147^{+10}_{-13}\right)\times10^{2}$    \\ 
\hline  
$A_\mathrm{ref}$  [$10^{-5}$]    &   $2.0\pm0.4$   &   $1.6^{+0.5}_{-0.4}$   &   $1.41^{+0.39}_{-0.15}$   &   $1.7^{+0.6}_{-0.5}$    \\
$\mathrm{Fe}/\mathrm{Fe}_\odot$   &   $3.6^{+0.9}_{-0.6}$   &   $6.0^{+2.5}_{-2.8}$   &   $3.9^{+0.8}_{-0.7}$   &   $3.9^{+1.0}_{-0.8}$ \\
$\xi$  $[\mathrm{erg}\,\mathrm{cm}\,\mathrm{s}^{-1}]$   &   $1440^{+310}_{-200}$   &   $1370^{+410}_{-230}$   &   $1910^{+190}_{-420}$   &   $1600^{+700}_{-400}$    \\
\hline  
$h$   &   $2.58^{+0.37}_{-0.30}$   &   $1.8^{+0.9}_{-0.4}$   &   $3.1\pm0.6$   &   $2.9^{+0.9}_{-0.7}$    \\ 
$a$   &   $0.921^{+0.077}_{-0.027}$   &   $0.995^{+0.004}_{-0.035}$   &   $0.9960^{+0.0020}_{-0.0837}$   &   $0.91^{+0.10}_{-0.07}$    \\ 
$i$ [deg]   &   $38.3^{+1.9}_{-2.0}$   &   $38.2^{+2.2}_{-3.3}$   &   $33.8^{+2.6}_{-2.3}$   &   $36^{+4}_{-6}$    \\ 
\hline  
$c_\mathrm{HEXTE}$   &   $0.844\pm0.006$   &   $0.823\pm0.006$   &   $0.834\pm0.004$   &   $0.826\pm0.005$    \\ 
$c_\mathrm{PN}$    &   $0.838^{+0.007}_{-0.005}$   &   $0.779^{+0.004}_{-0.005}$   &   $0.8132^{+0.0025}_{-0.0029}$   &   $0.813^{+0.004}_{-0.005}$    \\ 
$c_\mathrm{ISGRI}$   &   $0.923\pm0.009$   &   $0.916^{+0.010}_{-0.009}$   &   $0.903\pm0.008$   &   $0.905\pm0.008$    \\
\hline  
$s_\mathrm{gainshift}$   &   $1.0253^{+0.0018}_{-0.0029}$   &   $1.0199^{+0.0027}_{-0.0043}$   &   $1.0183^{+0.0010}_{-0.0009}$   &   $1.0210^{+0.0020}_{-0.0012}$    \\ 
\hline  
$\chi^{2}/\mathrm{dof}$ &  574.1/392&  410.4/319&  583.7/442&  405.8/331   \\ 
$\chi^{2}_\mathrm{red}$ &  1.46&  1.29&  1.32&  1.23   \\ 
\hline 
\end{tabular}

\end{table*}
Having shown that the best fit with a phenomenological model is
achieved when using a broken power-law emissivity profile with a steep
emissivity, we now turn to a physical model in which this emissivity
behavior is naturally produced. The motivation for this model comes
from the properties of the broad-band emission of Cyg X-1 that ranges
from radio to X-rays \citep[][and references
therein]{markoff:2005,wilms:2007,rahoui:2011}, and from the existence
of a short time-scale radio-X-ray flux correlation in \cyg
\citep{wilms:2006b, gleissner:2004}. X-ray emission from the source is
probably due to synchrotron self-Comptonization in a relativistic jet
that is launched along the rotational axis of the black hole
\citep{markoff:2005}. The ``lamp post'' geometry assumes the hard
X-ray radiation to be emitted from the base of this jet. As the source
is very close to the black hole at a height of only a few
$r_\mathrm{g}$, the light bending effects focus a large fraction of
the photons onto the accretion disk, producing the observed reflection
component. If the source is closer to the black hole, more photons
illuminate the disk, and less are left over to produce the continuum
component. With increasing height $h$, the opposite is true, creating
the anti-correlation between the continuum and the reflection
component, an effect that has already been observed in the AGN
MCG$-$6-30-15 \citep{miniutti:2004,miniutti:2006}. 

\begin{figure*}
  \centering
  \resizebox{17cm}{!}{\includegraphics{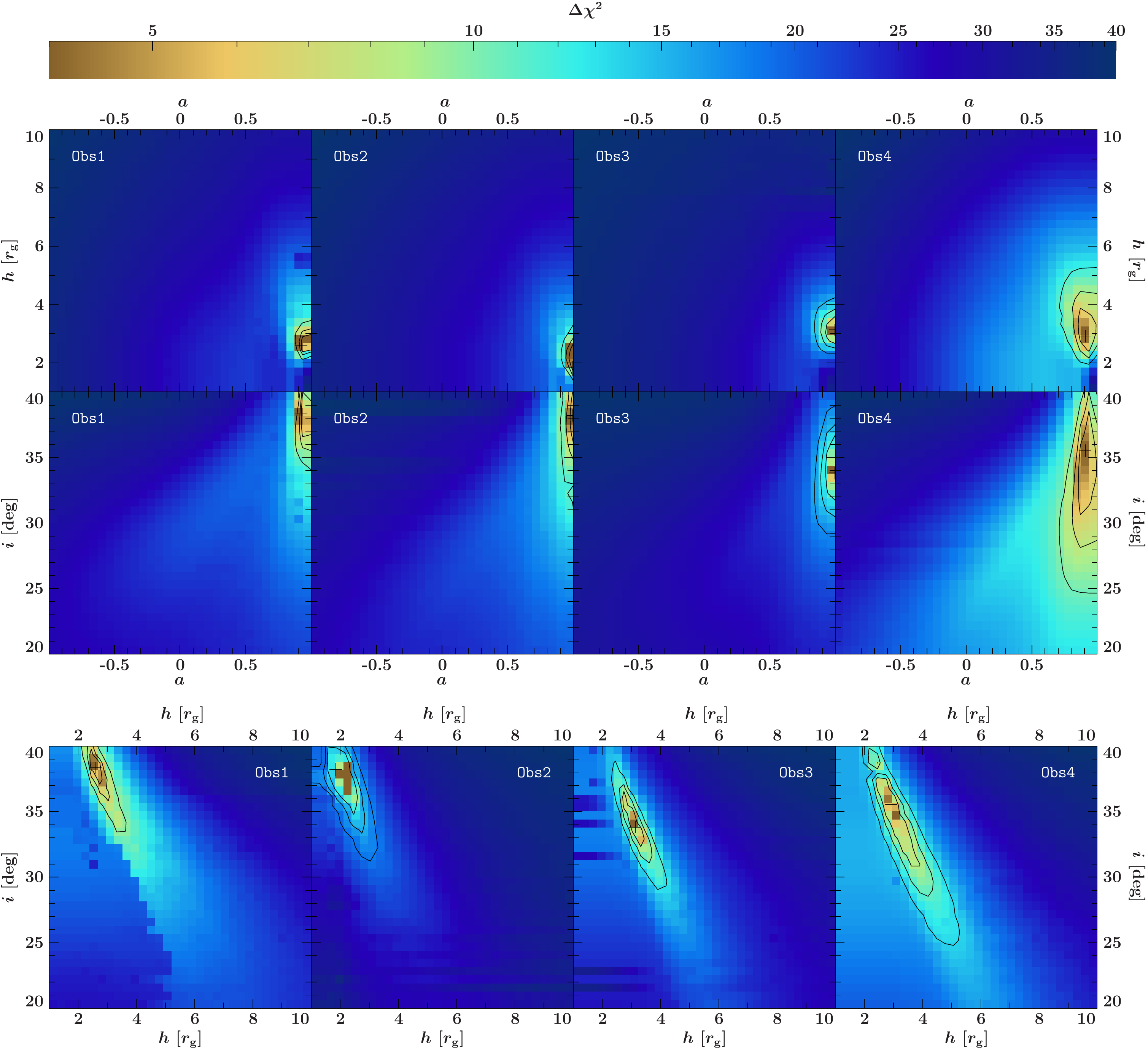}}
  \caption{Parameter correlations for the lamppost model (Model~4 in
    Table~\ref{tab:lp}). Shown are the $\chi^{2}$ significance
    contours for the two parameters of interest. First row: Jet base
    height $h$ in units of $GM/c^2$ versus the black hole spin $a$.
    Second row: Inner disk inclination $i$ expressed in degrees versus
    the spin $a$. The values clearly point towards a high spin value.
    Third row: Inner disk inclination $i$ versus the jet base height
    $h$. Shown are the $\chi^{2}$ significance contours for two
    parameters of interest based on $\Delta\chi^{2}=2.30$, 4.61, and
    9.21, i.e., 68\%, 90\%, and 99\% confidence, with respect to the
    best fit case. }
  \label{fig:cont_lp_incl}
\end{figure*}

For the spectral modeling we utilize the lamp post convolution model
\texttt{relconv\_lp} \citep{dauser:2013}. This model successfully
incorporates the relativistic effects that are imprinted in the
spectra due to the source's proximity to the black hole. In the model,
the height of the jet base, $h$, is a free parameter and expressed in
units of $r_\mathrm{g}$. In the model, light bending and aberration
effects lead to a ``focusing'' of the emission from the jet onto the
disk and directly produce a very steep emissivity profile for the
innermost few $r_\mathrm{g}$ of the disk \citep[see, e.g,][for
  examples of lamp post emissivity profiles]{dauser:2013}. To fit the
data, we use the continuum and the reflection model components as
deduced from the coronal geometry in
Sect.~\ref{subsec:coronal_geometry}. These spectral components are
expressed in the final model~4, which in ISIS notation, is given by
\begin{multline}
  N_\mathrm{ph}(E)=
  \mathrm{constant}\times(\mathrm{gabs}_{1}+\mathrm{gabs}_{2})\times \\
(\mathrm{cutoffpl}+\mathrm{diskbb}+\mathrm{egauss}+\mathrm{relconv\_lp}\otimes
\mathrm{reflionx})
\end{multline}
The best fit solution residuals for all four observations are shown in
Fig.~\ref{fig:residuals_lp}. They are hardly distinguishable from the
best fit solution using the coronal geometry (see
Fig.~\ref{fig:components}). The $\chi^{2}_\mathrm{red}$ values range
from 1.23 to 1.46 (with the same degrees of freedom). Inspecting the
\xmm residuals does not reveal any systematic variations, similar to
what was shown for the coronal geometry in
Fig.~\ref{fig:xmm_systematics}. In all four observations we find
extreme spin values ($a\ge0.9$). The jet height $h$ remains at very
low values, i.e., $h\sim3r_\mathrm{g}$. The low height of the jet base
indicates the presence of many primary photons irradiating the disk
and can therefore explain the significant amount of reflection
\citep{dauser:2013}. Although an anti-correlation between the jet
height $h$ and the inclination $i$ seems to be present (see
Fig.~\ref{fig:cont_lp_incl}), low values for both parameters are
required by the data. We note that the inclination $i\geq30^\circ$ is
somewhat higher than for the coronal geometry. The second row of
Fig.~\ref{fig:cont_lp_incl} shows that, irrespective of the
inclination $i$, the data consistently require high values of the
spin. Similarly, the $h$-$a$ dependence in the top row of the Figure
also demonstrates that a high spin value is required by the data. The
remaining relevant parameters are consistent with the values obtained
from the coronal geometry fits, i.e., $\mathrm{Fe/Fe}_{\sun}\sim4$,
$s_\mathrm{gainshift}\sim1.02$ and $E_\mathrm{fold}\sim170$\,keV.

As both geometric models lead to similar $\chi^2$ values and produce
consistent parameter values, we conclude that from our data it is
impossible to distinguish between the lamp post and the coronal
geometry. Similar model comparisons were also performed for \cyg using
\suzaku data \citep{nowak:2011} as well as for the galactic black hole
GX 339$-$4, for which \citet{markoff:2005} could show that their \rxte
data are equally well described when using both accretion flow
geometries \citep{markoff:2005}. Table~\ref{tab:lp} gives an overview
of the best fit values for the lamp post model.

\section{Discussion}
\label{sec:discuss}

\subsection{The spin of Cygnus~X-1}
Assuming that the inner edge of the accretion disk extends to the ISCO
allows us to interpret the changes of the intrinsically narrow Fe
K$\alpha$ emission line as the relativistic effects that arise due to
the closeness of the disk emission region to the black hole. Several
previous investigations are in favor of this assumption. For example,
\citet{reis:2010} have excluded truncation of the disk in the low-hard
state of 9 stellar mass black holes (including \cyg). We also refer to
the recent study of \cyg by \citet{miller:2012}, where the inner disk
radius in hard state \suzaku data was also found to be relatively
stable despite the changes in the power-law and the flux variations of
the radio component. \citet{miller:2012} find the spin in the range
$0.6\le a\le0.99$, which agrees well with the very high spin of \cyg
that we have derived here. Recent results coming from reflection
fitting \citep{fabian:2012}, continuum fitting \citep{gou:2011}, or
from both methods \citep{tomsick:2014}, strengthen the assumption and
confirm the high spin of \cyg.

\citet{gou:2011} analyzed soft state data from the Advanced Satellite
for Cosmology and Astrophysics, \rxte, and \chandra using the
continuum-fitting method \citep{zhang:1997}, finding $a\sim0.99$. The
results were examined by carrying out several tests (i.e., using
different reflection models and a change in inclination angle and
metallicity), all consistently showing a high spin value. These
authors derived the emissivity profile index of the disk to be
$\epsilon\sim2.8$, the convolution models \texttt{kerrconv} and
\texttt{kerrdisk} \citep{brenneman:2006}. The value agrees well with
our broken-power-law emissivity profile model for the outer disk
region ($\ge4 r_\mathrm{g}$) and with the low source height $h\sim3
r_\mathrm{g}$ for the jet base in the lamp post model. Our models also
require a high spin value for all physically meaningful values of the
relevant parameters such as $i$, $\epsilon$, $h$, and
$r_\mathrm{break}$, in agreement with the continuum-fitting results.

\citet{fabian:2012} measured the spin by fitting reflection models to
hard-state Suzaku data. Their result of $a=0.97^{+0.014}_{-0.02}$ is
based on fitting for an emissivity profile that goes as a broken-power
law, in which the steep value of $\epsilon\ge6.8$ is achieved for the
innermost region of $r_\mathrm{break}\le5.1r_\mathrm{g}$, while the
outer disk takes the standard thin disk emissivity profile value of
$\epsilon=3$. \citet{fabian:2012} note, however, that using a single
power-law with $\epsilon=3$ does produce an extreme spin value as
well. Using a single power-law emissivity profile we have previously
derived a similarly high value \citep{duro:2011}.

We are able to reproduce this result in all four observations for the
extended energy range used here (Sect.~\ref{sec:refl3}). The best
description, however, is achieved by using a broken power-law for the
emissivity profile. The spin is $a\sim0.9$, while the break radius is
at $\sim 4 r_\mathrm{g}$ with the outer region's emissivity index,
$\epsilon_\mathrm{out}$ fixed at 3. The profile is very steep for
observations Obs1 and Obs4, $\epsilon_\mathrm{in}\sim10$, for the
inner disk region. It less well constrained than in Obs2 and Obs3,
possibly owing to the flares and the short exposure time in Obs1 and
Obs4. The good description of the data by the broken power-law for the
emissivity profile is expected, specifically for compact emission
sources located near the central object \citep{dauser:2013}, as strong
gravitational effects require a steep profile for the inner disk
region \citep{wilkins:2011}.

In their analysis of the combined data of the 1--300\,keV simultaneous
Nuclear Spectroscopic Telescope Array (\nustar) and \suzaku data of
\cyg in the soft spectral state, \citet{tomsick:2014} model the
spectrum using both the reflection and the continuum fitting methods.
The broad energy coverage of these data enables them to constrain the
parameters of the thermal disk emission, the continuum, and the
reflection emission. They find the iron line to be very broad and
asymmetric with an extended red wing, i.e., to show clear signs of
relativistic influence on the intrinsic emission. The relativistic
reflection model components applied were the same as in our analysis
(\texttt{relconv}, \texttt{relconv\_lp}), but they modified the
intrinsic reflection model \texttt{reflionx} to accept higher
ionization levels than in its standard version used here.

An exploration of different models and parameter values showed that
the spin value from the accepted model fits was $a>0.83$
\citep{tomsick:2014}, which is consistent with our results. The
emissivity profile yields a high $\epsilon$ value for the inner disk
region, as also seen in our work. Interestingly, the inclination
inferred by \citet{tomsick:2014} implies a higher value of the
inclination than \citet{orosz:2011} found using optical spectroscopy.
We will return to the discussion of the inclination further below
(Sect.~\ref{sect:incl}).

We note that some earlier papers have claimed low spin values of
$a\sim0$ for Cyg~X-1 \citep[e.g.,][]{miller:2002, miller:2009}, which
are clearly inconsistent with our inferred spin for \cyg. Common to
these earlier models was that in addition to lacking precise
measurements of the distance, inclination, and mass of \cyg. They also
did not have the high energy coverage ($\ge10$\,keV) that is crucial
in order to constrain the underlying continuum and reflection features
properly \citep{ng:2010}. The current moderate abundance of the data
with high enough $\mathrm{S}/\mathrm{N}$ and broad energy coverage,
and the availability of more complex, yet more physically realistic
models, have helped to remove the inconsistency and achieve the
agreement on the high spin of \cyg. This is precisely what is
consistently concluded by the most recent results from the
disk-dominated state \citep{gou:2011,tomsick:2014}, the hard
intermediate state (this work), and the hard state
\citep{fabian:2012,miller:2012}.

\subsection{The accretion geometry of Cygnus~X-1}
It is difficult to distinguish between different proposed accretion
geometries. We examined the coronal geometry where we described the
disk emissivity profile with a single or broken power-law. As noted
before, both provide a good descriptions of the data, with a
preference for a break in the emissivity profile. Model fits with the
lamp post geometry lead to an equally good description of the data.
This similarity may point in the direction of a close connection
between the corona residing close to the black hole and the material
being collimated into a jet as suggested by \citet{markoff:2005},
building on earlier work on the accretion flow structure by, for
example, \citet{nowak:2002} and \citet{merloni:2002}. In this model,
hard photons are created in the jet base by synchrotron and
synchrotron self-Compton up-scattering. \citet{markoff:2005} compare
the model with thermal corona models, concluding that despite the high
statistics of RXTE spectra it is not possible to distinguish between
the jet model and thermal Comptonization. This result was later
confirmed by \citet{nowak:2011} with even higher signal to noise data
and with a better energy coverage.

\citet{miller:2012} propose several interesting properties that may
serve to put more constraints on the geometry of the system. Based on
twenty \suzaku hard state \cyg observations, the correlation between
the radio flux and the inner accretion disk temperature, as well as
the correlation with the reflected X-ray flux, suggest strong disk-jet
coupling. Using these findings, \citet{miller:2012} propose a system
where the corona/jet base is produced of the disk material that is
lifted by the magnetic forces, and (partly) accelerated by the
pressure of the reflected radiation \citep[for more details,
  see][]{miller:2012}. Such a geometry approximately identifies the
jet base with the corona, which could explain the similarities we
obtained from the coronal and the lamp post geometry. Ideally, this
could be further tested by simultaneous \pn and radio observations.

It is worthwhile to note that in the lamp post geometry, an
Fe~K$\alpha$ line that is very broad is already telling us that the
black hole is highly spinning and that the continuum source is compact
and is close to the black hole (i.e., with low height $h$). This is
due to the nature of the emission source, as concluded by
\citet{dauser:2013}. An elongated (and/or accelerated) jet produces
spectral lines that are narrow, and which are independent of the black
hole spin $a$. A similar line shape is achieved with a low spin and a
compact emission source, making the distinction from a high-spin
elongated jet case impossible. Only the combination of a high spin and
a low and compact jet produces a broad line that can be distinguished
from the other three cases \citep[see Fig.~10 of][]{dauser:2013}.

\subsection{The inclination of the inner disk}\label{sect:incl}

In all our fits the inferred inclination is close to
$i\sim30^\circ$. This agrees well with many previous studies that
consider different approaches to constrain the inclination of the \cyg
system. The lack of a significant dependence of the $\mathrm{H}\alpha$
emission strength on the orbital phase $\phi$ has led
\citet{sowers:1998} to constrain the inclination to $i\le55^{\circ}$.
Likewise, the lack of a two-sided jet detection suggested a low
inclination of the system \citep{stirling:2001}. The most
probable/realistic evolutionary models of \citet{ziolkowski:2005}
result in an inclination in the range of $28^{\circ}\le i
\le38^{\circ}$. More recently, based on optical measurements,
\citet{orosz:2011} adopt a model with eccentric orbit and asynchronous
rotation at periastron which produces an inclination of
$i=27\fdg1\pm0\fdg8$, and $i=28\fdg5\pm2\fdg2$ from their second best
model assuming synchronous rotation. Comparing these measurements of
the system's inclination with our measurement of the inclination of
the disk close to the black hole shows that in our intermediate/hard
state observations the inner disk and the orbital angular momentum are
in agreement. If we assume that the inner disk is aligned with the
black hole's spin, then we find that the spin axis and the orbital
momentum vector are aligned, as predicted, for example, by
\citet{bardeen:1975} and \citet{king:2005}.

There is, however, some discrepancy between our result and the high
precision inclination measurement made by \citet{tomsick:2014} who
used simultaneous \nustar and \suzaku observations in the soft state.
Their fits require an inclination that is at least $13^{\circ}$ higher
than the optically derived value of \citet{orosz:2011} while our data
agree with the optical inclination.

One possibility to explain this discrepancy is a systematic error in
the modeling approach. For example, we have shown in
Fig.~\ref{fig:contPL} that the steepness of the emissivity profile,
$\epsilon$, is strongly correlated with the inclination. Given the
high quality of the \nustar data, however, the systematic error is
likely to be much smaller than the discrepancy between the best fit
values for the inclination. In addition, we would also expect the spin
parameters to be different between the \nustar fits and our fits if a
systematic error was involed. This is not the case, and therefore it
is unlikely that systematic effects are to blame for the difference in
inclination.

If it is not due to systematics, then the difference in measured
inclinations must be a real physical effect. We first re-emphasize
that the inclination measured from the Fe K$\alpha$ line is the
inclination of the (thin) accretion disk close to the black hole,
while the inclination measured optically is the inclination of the
system as a whole. A difference between the inclination determined
from the Fe K$\alpha$ line and the optical inclination then implies
that the accretion disk may be warped in its inner regions. While it
is reasonable to assume that the angular momentum vector of the
accretion disk is aligned with the orbital angular momentum (i.e., the
angular momentum of the accreted material), the timescale for
accretion to align the black hole with the orbital angular momentum is
$10^6$--$10^8$ years (\citealt{steiner:12a} and references therein). A
full alignment between the disk and the orbit is therefore not
expected for a comparably young High Mass X-ray Binary. Furthermore,
starting with the pioneering work of \citet{bardeen:1975}, many
authors have shown that it is possible to produce a warp in the inner
regions of the disk \citep[e.g.,][]{pringle:96a,fragile:09a}.

Warped accretion disks often lead to long-term quasi-periodic
modulations which are observable in the X-ray lightcurves
\citep{clarkson:03b}, such as the 35\,d cycle of Her~X-1
\citep[e.g.,][]{staubert:09a} or the variable long-term period in
SMC~X-1 \citep{clarkson:03a,trowbridge:07a}. The well-known 150\,d or
300\,d quasi-periodicity of Cyg~X-1 has been associated with a similar
warp \citep[][and references
  therein]{priedhorsky:83a,kemp:83a,kemp:87a,brocksopp:99b,zdz:02a,benlloch:04a,zdz:2011}.

If the warp is due to some kind of radiative instability
\citep{pringle:96a,maloney:98a,clarkson:03a} then the inclination of
the inner disk produced by this warp could depend on the source
luminosity and/or the mass accretion rate. For example, as recently
shown by \citet{mckinney:13a}, thick disks with jets are much more
stable against warping than thin disks. According to these
simulations, the inner disk is therefore aligned with the orbital
angular momentum during the hard state and the measured inclination
should be the same for the binary orbit as for the inner accretion
disk. During the soft state, the inner disk is warped and the measured
disk and orbital inclinations can disagree. One can speculate whether
such a behavior is the reason for the difference of our results with
the soft state observations of \citet{tomsick:2014}.

As a caveat, however, and somewhat contrary to this hypothesis, we
note that the 150\,d period in Cyg~X-1 is strongest during the hard
state and not present during the soft state
\citep{benlloch:04a,zdz:2011}. As discussed by \citet{zdz:2011}, if
this period is due to a warped disk, then the cause for the warp is
probably not entirely due to irradiation effects. However, it might
also be that the 150\,d period is unrelated to the disk, and, for
example, due to jet precession without associated disk precession.

A further study of this behavior would require a detailed comparison
of the Fe K$\alpha$ line profiles measured in different states of
Cyg~X-1. A preliminary comparison of the NuSTAR profiles of
\citet{tomsick:2014} and our XMM profiles shows line profiles to be
different, independent of the continuum model used. The blue wing of
the NuSTAR line, which is most relevant for the determination of the
inclination, is more pronounced in these data, as expected for a
higher inclination. More recent hard state measurements with NuSTAR
have a less pronounced blue wing, which is more in line with the XMM
data presented here. A further comparison of all available high
signal-to-noise ratio line profiles of Cyg~X-1 is needed for a more
quantitative description. Such a comparison, however, is outside of
the scope of this work.

\section{Summary and conclusion}
\label{sec:summary}
In this work we presented the analysis of four \cyg broadband
(4--500\,keV) observations taken simultaneously with \xmm, \rxte, and
\integral. Our focus was on the broad Fe K$\alpha$ line, which is an
excellent tool for probing the strong gravity regime of the black
hole. We show that the black hole in \cyg is rapidly rotating, while
the system's inclination is at $i\sim30^\circ$. The \xmm\,\pn was
operated in the Modified Timing Mode that provided us with an
excellent $\mathrm{S}/\mathrm{N}$ coverage at the iron line region
around 6.4\,keV. The broad band data of RXTE-\pca, RXTE-\hexte and
INTEGRAL-\isgri were used to constrain the underlying continuum, which
is a crucial step as the shape and the strength of the Fe K$\alpha$
line depend strongly on the correct modeling of the continuum. Our
conclusions are the following:
\begin{enumerate}
\item \cyg is found in the hard intermediate state. The cutoff power
  law describes the continuum well, although a weak black body
  component with $T_\mathrm{BB}$ of a few hundred eV is additionally
  needed to describe the soft part of the spectra. The cutoff energy
  is at $E_\mathrm{fold}\sim170$\,keV.
\item We find a broad Fe K$\alpha$ line spectral feature at
  $E\sim6.4$\,keV. The broadening is due to the relativistic effects
  arising in the vicinity of the black hole, and the Doppler
  broadening due to the ionized reflector. The line cannot be
  described well with the simple modern relativistic Fe line models
  like \texttt{relline}. A full reflection model (\texttt{reflionx})
  convolved with a relativistic model (\texttt{relconv}) is needed to
  describe the reflection spectra.
\item The standard accretion flow geometry for black hole binaries in
  which the thermal electrons up-scatter the soft disk photons,
  describes the spectra well in all four observations. The result is a
  very rapidly spinning black hole with $a\sim0.9$ and an inclination
  of the inner, moderately ionized disk of $i\sim30^\circ$. A broken
  power-law for the emissivity profile of the disk provides best fits
  (Models~2 and~3), although similar results can be achieved
  with a Newtonian power-law emissivity profile of $r^{-3}$
  (Model~1).
\item The lamp post geometry, in which the base of the jet is the
  source for the continuum and the irradiation of the accretion disk,
  improves the fit slightly. For all four observations, the black hole
  parameters are similar to the coronal modeling in all four
  observations (Model~4). The distance of the jet base to the event
  horizon is small, resulting in a efficient irradiation of the inner
  accretion disk which produces significant disk reflection.
\end{enumerate}
The high signal to noise ratio of our data allows us to strongly
constrain the spin of the black hole. We demonstrated that the spin
values inferred from our four observations are consistent, thus
reducing the probability that our result is accidental. In addition,
our result for the spin is also consistent with spin measurements
found from line fitting with other instruments, such as Suzaku/NuSTAR,
and with measurements performed during different source states. The
line fitting method therefore gives consistent results irrespective of
the accretion flow geometry. Our result is also consistent with the
value obtained from continuum fitting, i.e., with a completely
independent ansatz.

The next step in the endeavour of understanding Cyg~X-1 is to find
more ways to constrain the different accretion flow geometries, which
cannot be distinguished from spectral fitting alone. The difference in
the inner disk inclinations between the recent soft state
Suzaku/NuSTAR measurement of \citet{tomsick:2014}
and our results can be interpreted as an indication for a direct
change in accretion geometry that has to be followed up by a more
comprehensive survey of the Fe K$\alpha$ line shapes in Cyg~X-1.
Together with improvements in the available physical models we hope
that further constraints on the accretion flow geometry in this
crucial system will soon be obtained.

\begin{acknowledgements}
  We thank Norbert Schartel and the \xmm operations team for agreeing
  to perform observations in a new and untested mode, and Maria
  D\'iaz-Trigo for many useful discussions on CTE and pile up effects
  in the EPIC-pn camera. This work was partly supported by the
  European Commission under contract ITN~215212 ``Black Hole
  Universe'' and by the Bundesministerium f\"ur Wirtschaft und
  Technologie under Deutsches Zentrum f\"ur Luft- und Raumfahrt grants
  50\,OR\,0701, 50\,OR\,1007, and 50\,OR\,1113. Further support for
  this work was provided by NASA through the Smithsonian Astrophysical
  Observatory (SAO) contract SV3-73016 to MIT for Support of the
  Chandra X-Ray Center (CXC) and Science Instruments. CXC is operated
  by SAO for and on behalf of NASA under contract NAS8-03060. JAT
  acknowledges partial support from NASA Astrophysics Data Analysis
  Program grant NNX13AE98G. We acknowledge the support by the DFG
  Cluster of Excellence ``Origin and Structure of the Universe''. We
  are grateful for the support of M. Cadolle Bel through the
  Computational Center for Particle and Astrophysics (C2PAP). JR
  acknowledges funding support from the French National Research
  Agency, CHAOS project ANR-12-BS05-0009
  (\url{http://www.chaos-project.fr}). This research has made use of
  ISIS functions provided by ECAP/Remeis observatory and MIT
  (\url{http://www.sternwarte.uni-erlangen.de/isis/}). We thank John
  E. Davis for development of the \texttt{SLXfig} package that was
  used to create the figures throughout this paper and Sasha
  Tchekhovskoy for useful discussions on accretion disk warping. This
  paper is based on observations obtained with \xmm, an ESA science
  mission with instruments and contributions directly funded by ESA
  member states and NASA.
\end{acknowledgements}

\appendix

\section{Calibration of the Modified Timing Mode} \label{app:A}

In this and the following sections we discuss the \xmm EPIC-pn
Modified Timing Mode and its calibration. We start with the motivation
for the mode, then discuss the principle behind the calibration
necessary, show tests of the calibration, and conclude with some
additional caveats.

\subsection{Motivation}
The telemetry bandwidth allocated to the EPIC-pn is restricted to
$16\,\mathrm{kbit}\,\mathrm{s}^{-1}$ \citep{kendziorra:2004}. Because
of the large data rates produced by the instrument when observing
bright sources, it is not trivial to telemeter all information to
ground using the available data modes.

The standard data mode of the EPIC-pn for very bright sources is the
Burst Mode. In this mode the instrument serves as a ``grey filter''
and only 3\% of the total exposure time is ``live'', i.e., only events
taken during 3\% of the exposure are telemetered to ground.

In contrast, the \pn Timing Mode mode has a live time of 99.5\% that
is achieved by a continuous read out of the CCD. The problem of this
mode is the telemetry-induced limit of
$250\,\mathrm{events}\,\mathrm{s}^{-1}$ on the maximum average count
rate. If the average count rate surpasses this threshold, the on board
data buffers can overflow. The instrument reacts to these buffer
overflows by inserting ``telemetry gaps'', i.e., by throwing away all
science data for a certain time until the full data buffers have
emptied out. Outside of telemetry gaps the full data (spectral and
timing information) are available. From ground, no control can be
excerted over when such gaps occur in the data stream. When the goal
is to study, for example, the time variability of the observed source,
this is highly problematic. We stress that here and in the following
we talk about \emph{average} count rates. Short term flux variations
are compensated for by storing the higher count rate data in the on
board buffers.

A first step to minimize the probability of telemetry gaps is to
provide the EPIC-pn with as much telemetry as is possible. As
suggested by \citet{kendziorra:2004}, the total telemetry available to
the EPIC-pn camera can be increased by switching off the EPIC-MOS
cameras (or at least the central MOS camera). None of the EPIC-MOS
data modes are useable for very bright sources, so that this approach
is approriate if no data from sources close to the main target is
required. The EPIC-pn bandwidth inceases so to
$40\,\mathrm{kbit}\,\mathrm{s}^{-1}$, corresponding to a count rate
limit of $1050\,\mathrm{counts}\,\mathrm{s}^{-1}$.

If this higher count rate limit is not yet sufficient and if the
scientific purpose of an observation does not require the full energy
band available -- as is the case, for example, for studies of
broadened Fe K$\alpha$ lines -- then the count rate limit can be
further increased by only telemetering events that are required for
the desired scientific analysis. This is achieved by changing the
lower energy threshold (LET) of the on-board electronics. The LET is a
programmable parameter of the EPIC-pn which sets the threshold energy
above which an event is telemetered to ground. If the energy deposited
in a pn-pixel is below the LET, such an event is discarded on board.
The default value of the LET is 150\,eV. The original purpose of the
LET was to avoid wasting telemetry for thermal noise produced in the
detector. However, it can also be used to exclude soft X-ray data from
being telemetered. In the case of the observations studied here, the
LET was set to 2.8\,keV. As discussed in the next section, the change
of the LET requires the creation of a new response matrix for the data
analysis.
       
\subsection{Calibration of the Modified Timing Mode}
X-ray CCDs measure the photon energy of incoming X-rays by measuring
the total charge deposited in the active parts of the detector
silicon. As the size of the individual pixels of the EPIC-pn is
comparable to the size of the charge cloud in the detector device, for
$\sim$25\% of all incoming X-ray photons charge is detected in more
than one pixel. Depending on the number of pixels in which charge is
detected, these ``split events'' are called single, double, triple, or
quadruple events.

In contrast to other X-ray astronomical CCDs such as those on Chandra
or Suzaku, EPIC-pn does not perform a recombination of double to
quadruple events on board, i.e., no ``event grading'' is done on
board. Rather, the information in all single pixels in which energy is
deposited (the ``split partners'') and for which that deposited energy
is above the LET are telemetered to ground. Based on this event
information, the photon energy is then reconstructed in the SAS by
adding the energies of all split partners.

A consequence of this approach of reconstructing split events only on
ground is that for all split events for which the energy of one or
more of the split partners is below the LET, the corresponding energy
is lost. This effect leads to a degradation of the energy resolution
of double, triple, or quadruple events compared to the single events.
Events where split partners are below the LET are also assigned the
wrong grade. For example, a double event with a split partner below
the LET would be erroneously classified as a single event.

The increase of the LET means that this energy degradation has to be
taken into account, i.e., a new detector response matrix file (RMF) is
required for the Modified Timing Mode observations. The most obvious
effect of loosing these split partners is that higher energy photons
appear to be softer as the events are wrongly recorded as single
events (see Figs.~\ref{fig:thick_filter}--\ref{fig:thin_filter}).

Thankfully, the response matrix for the Modified Timing Mode can be
constructed using archival observations made with standard Timing Mode
data, without requiring any new calibration observations
\citep[see][for an earlier description of the calibration
  procedure]{fritz:2008}: In the standard approach of X-ray
astronomical data analysis the response matrix $R_{j,k}$ corresponds
to the probability of observing a photon in the energy bin $E_{k-1}\le
E < E_k$ in the detector channel $j$. The counts detected in channel
$i$ of a Modified Timing Mode observation, $c_i$, can then be written
as
\begin{equation}\label{eq:rmf}
c_i = \sum_j P_{i,j} \sum_k R_{j,k} F(E_k) \Delta E_k
\end{equation}
where $F(E_k) \Delta E_k$ is the photon flux in the energy band
corresponding to the $k$th energy bin of the original response matrix
$R_{j,k}$ and where $P_{i,j}$ describes the probability that a photon
that would have ended up in energy channel $j$ in the standard Timing
Mode is observed in channel $i$ of the spectrum measured in the
Modified Timing Mode. Equation~\eqref{eq:rmf} therefore says that if
we know the redistribution matrix $P_{i,j}$ from the Timing Mode to
the Modified Timing Mode, we can construct the detector response
matrix of the Modified Timing Mode by multiplying the Timing Mode
response matrix with $P_{i,j}$.

It is possible to construct $P_{i,j}$ from archival Timing Mode
observations: For a set of Timing Mode observations we construct fake
Modified Timing Mode observations by filtering out all raw events with
energies below the increased LET of the Modified Timing Mode and then
produce new event files using the standard SAS tools. For each event
in the Timing Mode event file we can then look up the new
reconstructed energy in the fake Modified Timing Mode event file and
thus, by using all suitable Timing Mode observations, calculate the
probabilities $P_{i,j}$ to high precision. In practice it turns out
that the events in the fake Modified Timing Mode observation can
almost perfectly be associated with events in the original Timing Mode
observation by comparing the photon arrival times and the location of
the event on the detector. A set of computer routines in the
Interactive Data Language (IDL) to perform this calculation is
available upon request and documented by \citet{fritz:2008}.

A similar procedure can also be used for the construction of any
response matrix for a new detector working on a similar principle,
i.e., it is sufficient to calibrate a detector only for one low value
of the LTE. All other calibration files can then be constructed using
the appropriate $P_{i,j}$ matrices.

In order to construct $P_{i,j}$ for the observations studied here we
looked at all available Timing Mode observations. We then discarded
all those observations where deviations between the theoretical,
energy dependent distribution of multiple events and the observation
was seen, since these observations are likely piled up (i.e., more
than one photon hit the same or adjacent pixels during the one CCD
exposure). We also discarded all observations that could have been
contaminated by optical photons as electrons generated in the CCD by
the absorption of an optical photon shift the energy scale of the
detector (``optical loading'') by
$n_{\mathrm{op}}\times3.68\,\mathrm{eV}$, where $n_\mathrm{op}$ is the
number of optically generated electrons. This shift distorts
$P_{i,j}$.

During the course of this work a second important effect was noticed
by the XMM calibration team at ESA's European Space Astronomy Centre
(ESAC), Villafranca. This effect, ``X-ray loading'' affected a large
number of earlier Timing Mode observations of bright X-ray sources.
The reason for X-ray loading is as follows: Before each EPIC-pn
observation a short exposure is taken to measure the baseline noise in
the detector. This ``dark frame'' is then subtracted off on board
before measuring the energy deposited in each pixel. Unfortunately,
for bright X-ray sources the dark frame can be affected by X-ray
photons if the detector is not shielded from X-rays. X-ray loading in
Timing Mode can thus affect one or more columns of the CCD. It
effectively leads to a change of the energy assigned to all events
hitting a CCD row. This in turns leads to pattern migration from
higher to lower pattern types and finally to a change in the energy
spectrum. Using the closed filter in place during the offset map
exposures removes the possibility for the presence of X-ray loading.
For the Modified Timing Mode observations the offset was determined
with the filter in the closed position and therefore our observations
are not affected. This is not true for many of the archival Timing
Mode Observations, however. To minimize its effect on our analysis, we
excluded all observations of very bright sources in our calculation of
$P_{i,j}$.

In a final step, the response matrix for the Modified Timing Mode is
produced by a simple matrix multiplication with the Timing Mode
response matrix determined for the optical blocking filter used in the
observation. The response matrices used here are available through the
authors as well as through the \xmm calibration database available
from ESAC.
  
\subsection{Validation of the Modified Timing Mode Calibration}

\begin{figure}
  \centering \resizebox{\hsize}{!}{\includegraphics{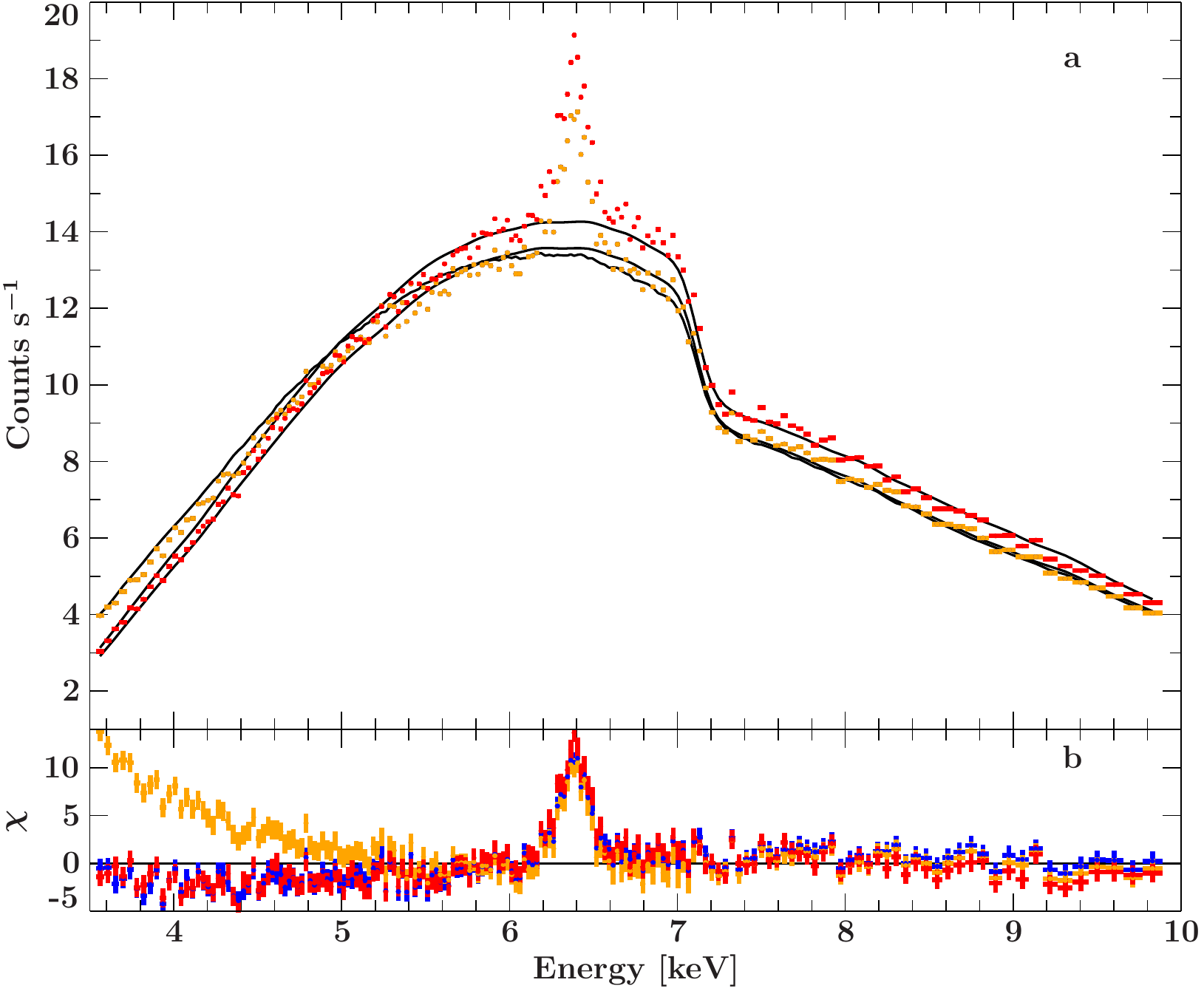}}
  \caption{Top: Observation of Vela~X-1 (\xmm ObsID 0111030101) made
    with the standard Timing Mode (red data points) and how the same
    observation would have looked like in the Modified Timing Mode
    (orange data points). Bottom: Residuals of fitting these data with
    an absorbed power-law (red). Applying the same model and the
    standard Timing Mode matrix produces the residuals shown in
    orange, illustrating the softening. The residuals of a fit with
    the Modified Timing Mode response matrix (blue) agree with the
    Timing Mode residuals.}
  \label{fig:thick_filter}
\end{figure}

\begin{figure}
  \centering
  \resizebox{\hsize}{!}{\includegraphics{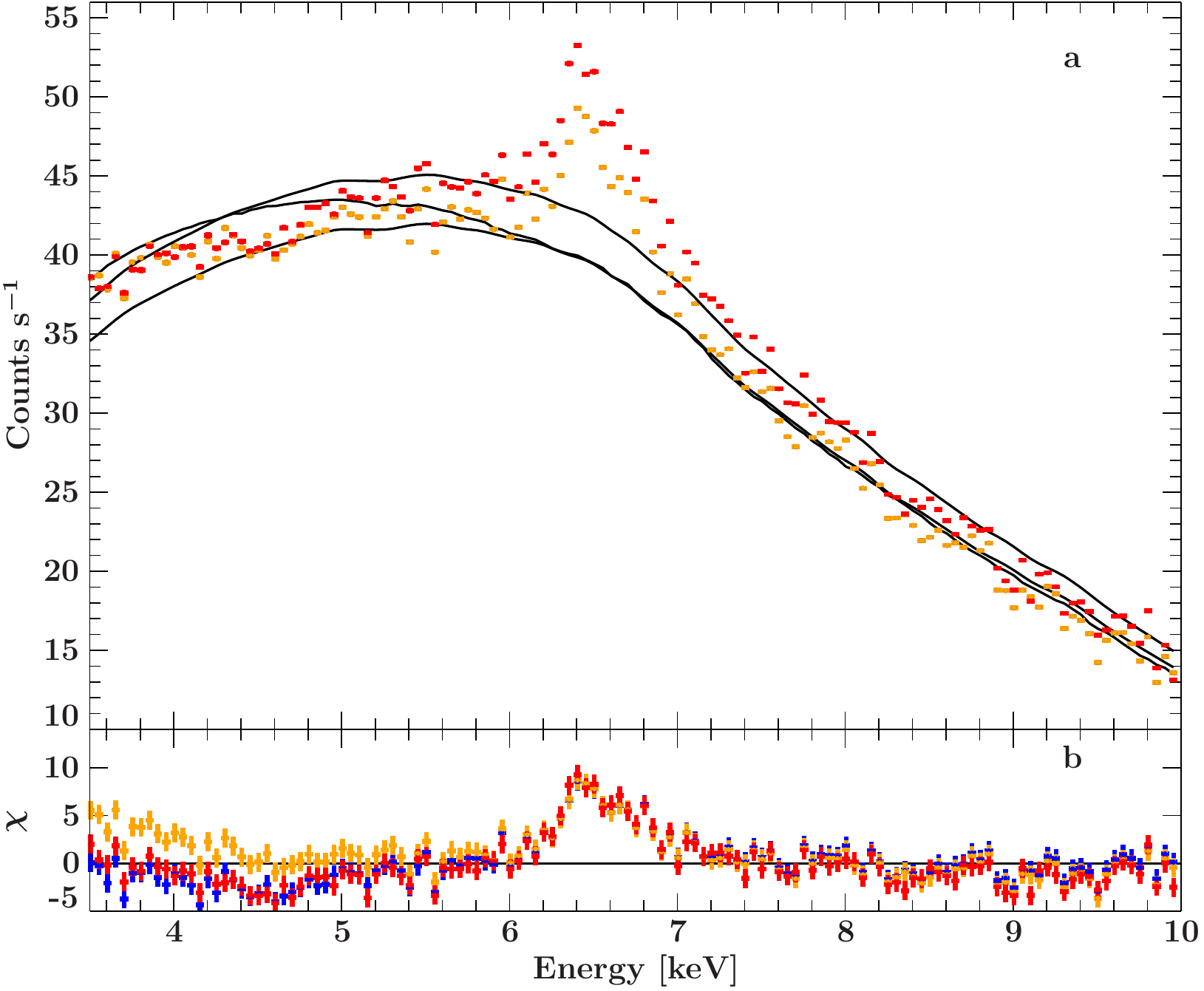}}
  \caption{Same as Fig.~\ref{fig:thick_filter}, but for an observation
    of Her X-1 (ObsID 0134120101) using the medium filter.}
  \label{fig:medium_filter}
\end{figure}

\begin{figure}
  \centering \resizebox{\hsize}{!}{\includegraphics{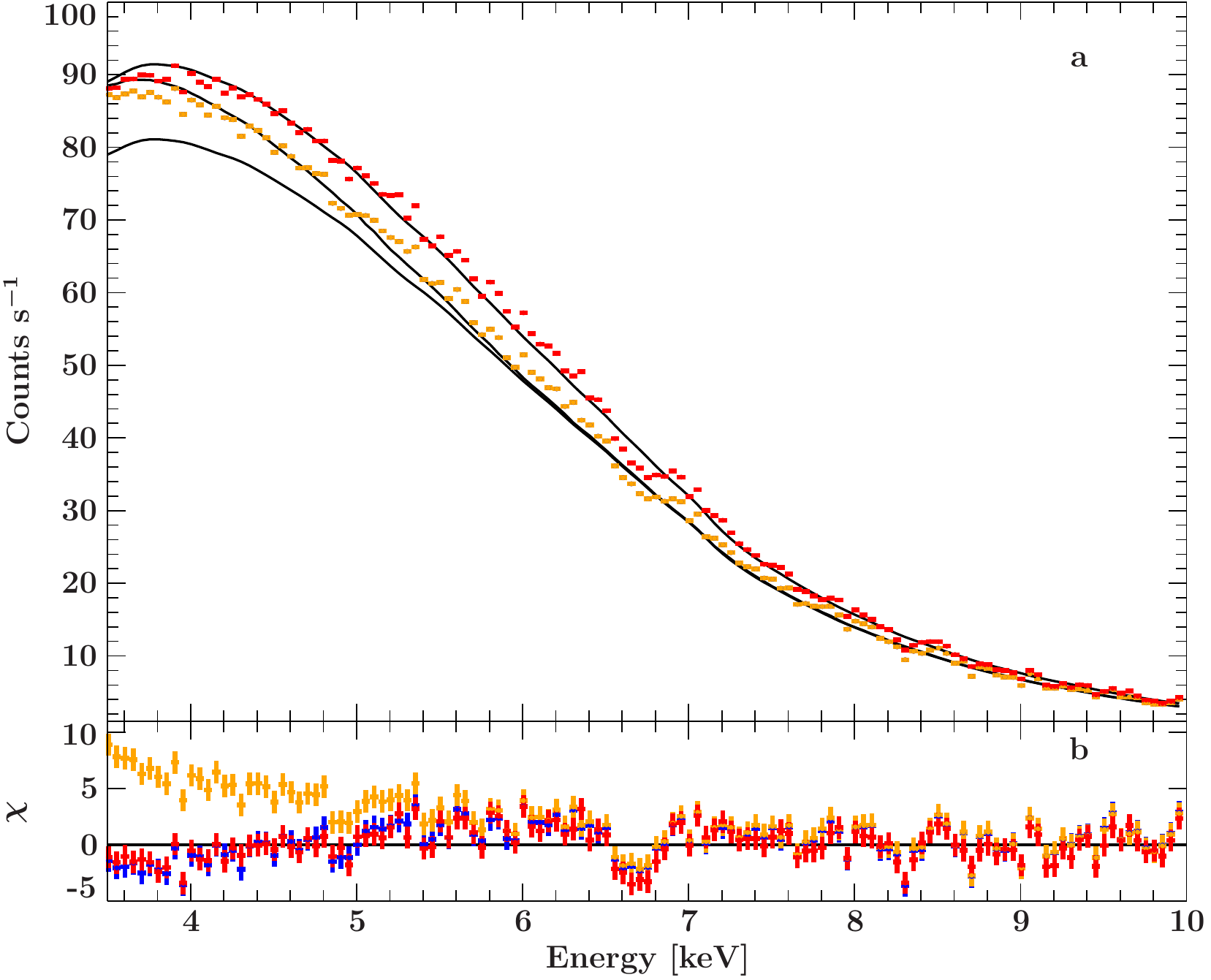}}
  \caption{Same as Fig.~\ref{fig:thick_filter}, but for an observation
    of Circinus X-1 (ObsID 0307420601) using the thin filter.}
  \label{fig:thin_filter}
\end{figure}

The Modified Timing Mode response matrix used in this paper was
obtained based on 82 thick filter observations (see
Table~\ref{tab:thick_obs}) and with a LET of 2.8\,keV. Because the
source was so bright, the inner three columns of the point spread
function were discarded in the actual observations of Cyg X-1 used in
our analysis. This effect was also taken into account in the
construction of the response matrix.

To check how well the new response matrix can reproduce spectral
features and to illustrate the effects of the increased LET we
generated fake Modified Timing Mode observations of a set of bright
sources using the tick, medium, and thin filters of the EPIC-pn camera
(Fig.~\ref{fig:thick_filter}--\ref{fig:thin_filter}).
Figure~\ref{fig:thick_filter} shows the data points of the Timing Mode
observation of the Galactic High Mass X-ray Binary Vela X-1 (\xmm
ObsID 0111030101; red data points). To illustrate the effects of the
response matrices, we fit a simple absorbed power-law to these data
(Fig.~\ref{fig:thick_filter}, bottom). Removing all raw events below
the LTE of 2.8\,keV, and reconstructing again the event list results
in the orange data points. Note that the loss of split partners below
the LTE leads to a softening of the spectrum below about 5\,keV. This
is well illustrated in the residuals of the absorbed power-law fit if
we apply the normal Timing Mode response matrix (orange data points).
Using the Modified Timing Mode response matrix, on the other hand,
yields residuals (and best fit values) which are identical to those of
the Timing Mode observation. This illustrates that $P_{i,j}$ was not
affected significantly by statistical noise.

Similar calculations are shown in Figs.~\ref{fig:thin_filter}
and~\ref{fig:medium_filter}, which use the thin and medium filters,
respectively. Even though no Modified Timing Mode observations exist
which employ these filters, these measurements indicate that the
calibration for these two filters is of similar quality as that for
the thick filter.

\begin{table}
  \caption{\xmm ObsIDs for thick filter observations used
    to produce the modified response matrix for the respective
    filter.    \label{tab:thick_obs}}  

\centering
\begin{tabular}{cccc}
\hline
\hline
0084110201 & 0084020401 & 0084020501 & 0084020601 \\ 
0099280101 & 0111030101 & 0111040101 & 0111370101 \\ 
0111400101 & 0111410101 & 0111470101 & 0111470201 \\
\hline
0111470301 & 0111500101 & 0112900101 & 0122330801 \\ 
0124930301 & 0124930601 & 0133120101 & 0134721501 \\
0149550401 & 0149810101 & 0153750401 & 0157960101 \\ 
\hline
0158971301 & 0159360401 & 0159361101 & 0160960201 \\ 
0160960301 & 0160961001 & 0160961201 & 0160961401 \\ 
0201220101 & 0203500201 & 0206320101 & 0206670101 \\
\hline
0206670201 & 0212480501 & 0303220201 & 0305240101 \\
0305240201 & 0305240301 & 0305240401 & 0305240501 \\
0400550201 & 0402300201 & 0405510701 & 0410580401 \\
\hline
0410580501 & 0412590201 & 0412590601 & 0412591101 \\
0412591501 & 0412592001 & 0412592501 & 0506110101 \\ 
\hline
\hline
\end{tabular}

\end{table}
  
\begin{table}
  \caption{\xmm ObsIDs for medium filter observations used
    to produce the modified response matrix for the respective
    filter.}    \label{tab:medium_obs}  
  \centering
\begin{tabular}{cccc}
\hline
\hline
0031740101 & 0056030101 & 0061140101 & 0061140201 \\ 
0085290301 & 0085680501 & 0085680601 & 0087350101 \\ 
0087350601 & 0087350801 & 0087350901 & 0092820101 \\ 
0092820201 & 0092820301 & 0092820801 & 0092821201 \\ 
\hline
0109090101 & 0111060101 & 0111061201 & 0111061301 \\ 
0111061401 & 0111061501 & 0111061601 & 0111061701 \\ 
0111061801 & 0111062101 & 0111062301 & 0111062501 \\ 
0111062601 & 0111230101 & 0111310201 & 0111390101 \\ 
\hline
0111390301 & 0111400101 & 0111410101 & 0111470101 \\ 
0111470201 & 0111470301 & 0111490101 & 0111490401 \\ 
0111500101 & 0113020201 & 0113050101 & 0113050201 \\ 
0113050701 & 0124930301 & 0128120401 & 0134120101 \\ 
\hline
0136541001 & 0137550301 & 0148220201 & 0153950301 \\ 
0158971201 & 0164570301 & 0165360101 & 0165360201 \\ 
0165360401 & 0202400701 & 0206670101 & 0206670201 \\ 
0210681401 & 0210682801 & 0303250201 & 0311590901 \\ 
\hline
0312590101 & 0402330301 & 0402330501 & 0402330601 \\ 
0405510201 & 0405510701 & 0406620301 & 0406700201 \\ 
0411080701 & 0413180201 & 0413180301 & 0500350301 \\ 
0500350401 & 0502211101 & 0503190301 & 0503190401 \\ 
\hline
0506291201 & 0510610201 & 0552270501 & 0604030101 \\ 
0605370101 & 0605610201 & 0610000901 & 0653110101 \\ 
\hline
\hline
\end{tabular}

\end{table}
  
\begin{table}
  \caption{\xmm ObsIDs for thin filter observations used
    to produce the modified response matrix for the respective
    filter.}    \label{tab:thin_obs}  
  \centering
\begin{tabular}{cccc}
\hline
\hline
0023940401 & 0036140201 & 0060740101 & 0060740901 \\ 
0064940101 & 0073140201 & 0073140301 & 0073140501 \\ 
0074140101 & 0074140201 & 0090340101 & 0090340201 \\ 
0090340601 & 0094520201 & 0094520301 & 0111100101 \\ 
\hline
0111100201 & 0111100301 & 0112200101 & 0112320101 \\ 
0112320201 & 0112320601 & 0112440101 & 0144900101 \\ 
0146870401 & 0146870501 & 0146871001 & 0146871501 \\ 
0148590201 & 0148590301 & 0148590401 & 0148590501 \\ 
\hline
0148590701 & 0148590801 & 0148590901 & 0150390101 \\ 
0150390301 & 0150495601 & 0150498701 & 0151412201 \\ 
0151412401 & 0153090101 & 0154750301 & 0201590101 \\ 
0205920501 & 0205920601 & 0302180101 & 0304080301 \\ 
\hline
0304080401 & 0304080601 & 0304720101 & 0307400101 \\ 
0307420401 & 0307420501 & 0307420601 & 0401390101 \\ 
0403530201 & 0405510301 & 0405510401 & 0405510501 \\ 
0405520301 & 0410180101 & 0410180201 & 0412981501 \\ 
\hline
0412981601 & 0502030101 & 0504370401 & 0505480201 \\ 
0505480301 & 0505480501 & 0555410101 & 0560180201 \\ 
0560180601 & 0605160101 & 0610000701 & 0652730101 \\ 
\hline
\hline
\end{tabular}

\end{table}
  
\section{Bright Sources and the Charge Transfer Inefficiency
  (CTI)} \label{app:C}

Irrespective of the data mode bright source observations with \xmm are
affected by a further instrumental effect, namely the influence that
the source photons have on the Charge Transfer Inefficiency of the
detector. Like all X-ray astronomical CCDs the \pn suffers from
contamination of the detector silicon (in the case of the EPIC-pn the
contamination is due to titanium). This contamination causes traps in
the silicon bulk which lead to a significant loss of the total
generated signal charge when they are transferred to their closest
read-out node. The effect is primarily dependent on the generated
signal charge, and therefore the charge transfer efficiency (CTE)
strongly correlates with the surface brightness on the CCD
\citep{kendziorra:1997}. Calibration measurements for fainter
observations allow us to model the CTE. Taking the measured source flux
into account it is then possible to reconstruct the original event
energy from its measured (CTE-affected) value. Using SAS v10 this
correction works well for faint sources\footnote{Using the newer
  version SASv11 does not improve calibration on observations used in
  this paper. The gain-shift correction in the model is still
  necessary to get a reasonable fit. See \citet{pintore:2014} and
  XMM-CAL-SRN-0083.}, but for sources as bright as Cyg~X-1 the CTE
correction is highly nonlinear and cannot be modeled. Applying the
standard CTE correction to such data inevitably leads to an
overcorrection and finally wrong parameters describing the observed
object. This forces us to apply a different approach using
simultaneous or similar flux data from other X-ray missions.

\begin{figure}
  \centering
  \resizebox{\hsize}{!}{\includegraphics{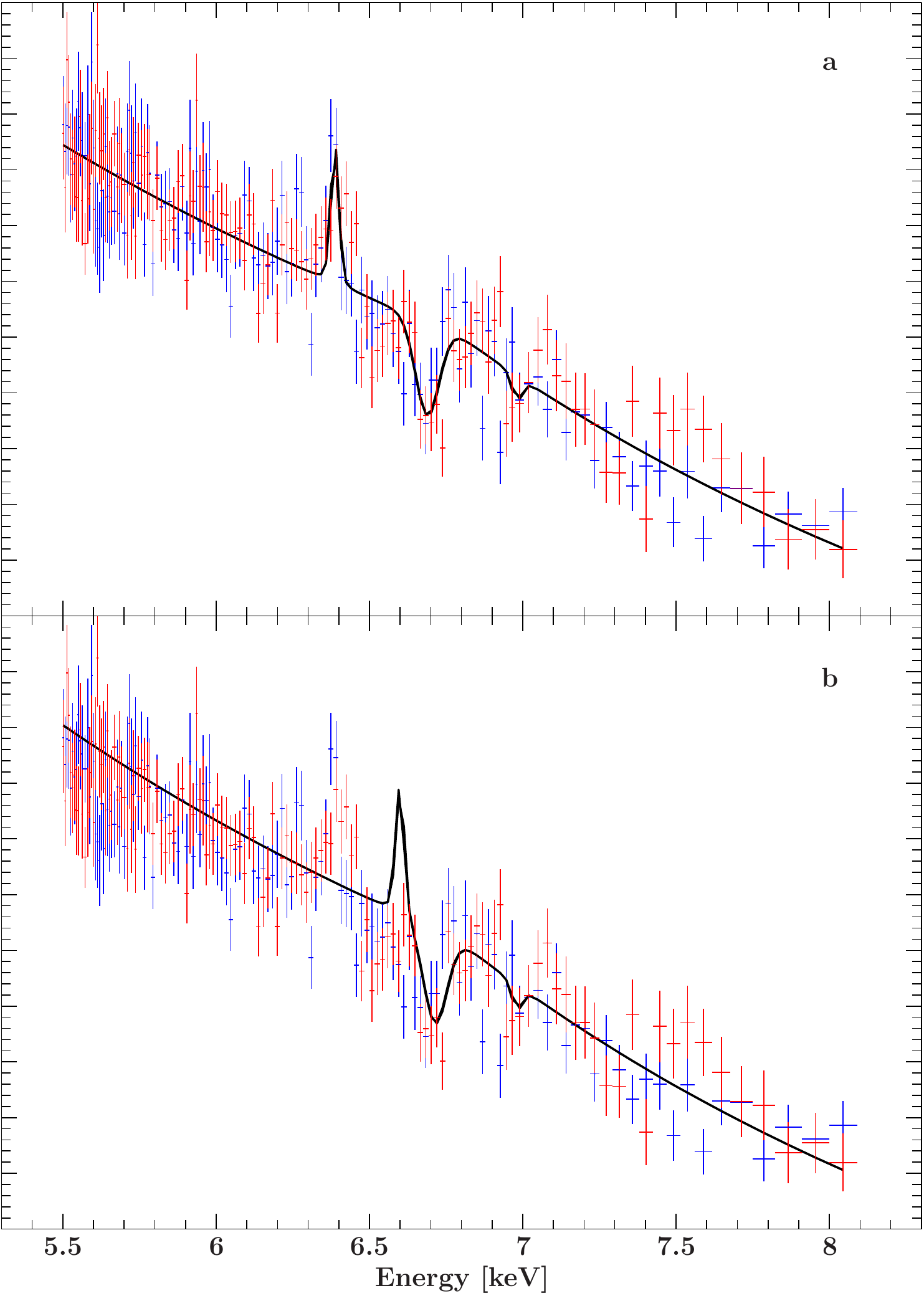}}
  \caption{Gain shift in the \pn. (a) Fit to the \chandra data. The
    model is composed of a broken power-law, an emission line at 6.4
    keV and two absorption lines for \ion{Fe}{xxv} and \ion{Fe}{xxvi}.
    The two HEG datasets are presented in different colors. (b)
    Applying a model taken from the fit to the \pn data and applied to
    \chandra data reveals the shift towards higher energies. Applying
    a gain-shift of $\sim$2\% to the \pn data corrects the fit and
    re-produces the good fit in \texttt{a)}. The line widths were kept
    fixed for plotting purposes.}
  \label{fig:chandraGain}
\end{figure}

\begin{figure}
  \centering \resizebox{\hsize}{!}{\includegraphics{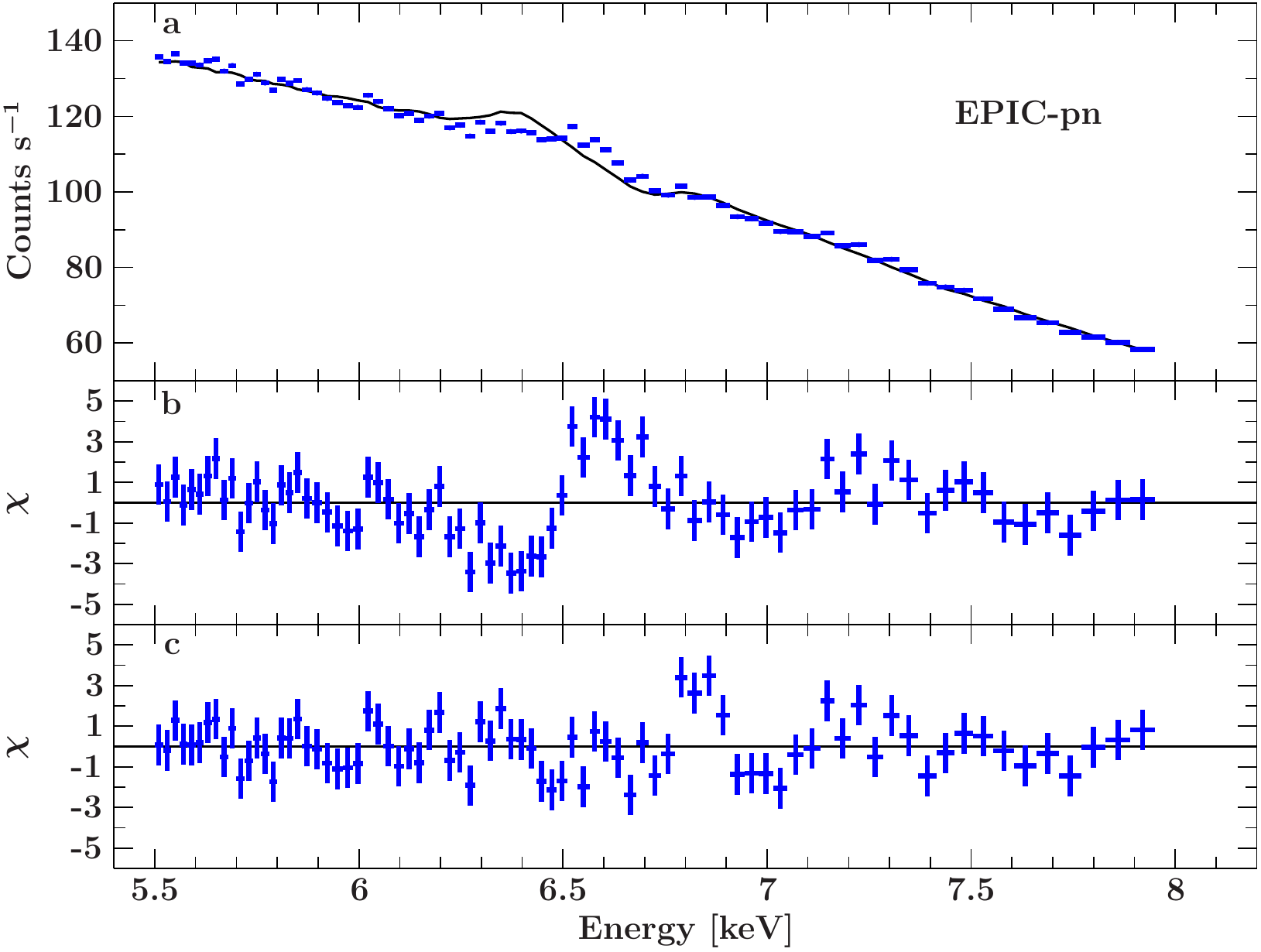}}
  \caption{Gain shift correction of the EPIC-pn data. (a) Fit to the
    \pn data of an absorbed power-law with narrow emission and
    absorption components from \chandra. (b) Large residuals remain
    around 6.5\,keV. (c) Applying a gain-shift of $\sim$2\% to the \pn
    data improves the fit considerably.}
  \label{fig:xmmGain}
\end{figure}

\begin{table}
  \caption{Best fit to the absorption lines seen in the \chandra data.
    $\tau$ describes the line depth and $\sigma$ its
    width.}  \label{tab:chandra_fit}  
  \centering
\renewcommand{\arraystretch}{1.3}
\begin{tabular}{lc}
\hline
\hline 
$E_\mathrm{\ion{Fe}{xxv}\ K\alpha}$  [keV]  &$6.689^{+0.010}_{-0.012}$\\
$\tau$ & $0.012^{+0.005}_{-0.004}$ \\
$\sigma$ [keV]& $0.028^{+0.020}_{-0.012}$\\
\hline
$E_\mathrm{\ion{Fe}{xxvi}\ K\alpha}$  [keV]   & $6.954^{+0.247}_{-0.004}$\\
$\tau$ $[10^{-2}]$   & $0.17^{+0.35}_{-0.17}$\\
$\sigma$ [keV] & $\le10$\\
\hline
$\chi_\mathrm{red}$/dof &299/263 \\
$\chi_\mathrm{red}^2$ & 1.13 \\
\hline
\hline
\end{tabular}
\end{table}
  
\begin{figure}
  \centering \resizebox{\hsize}{!}{\includegraphics{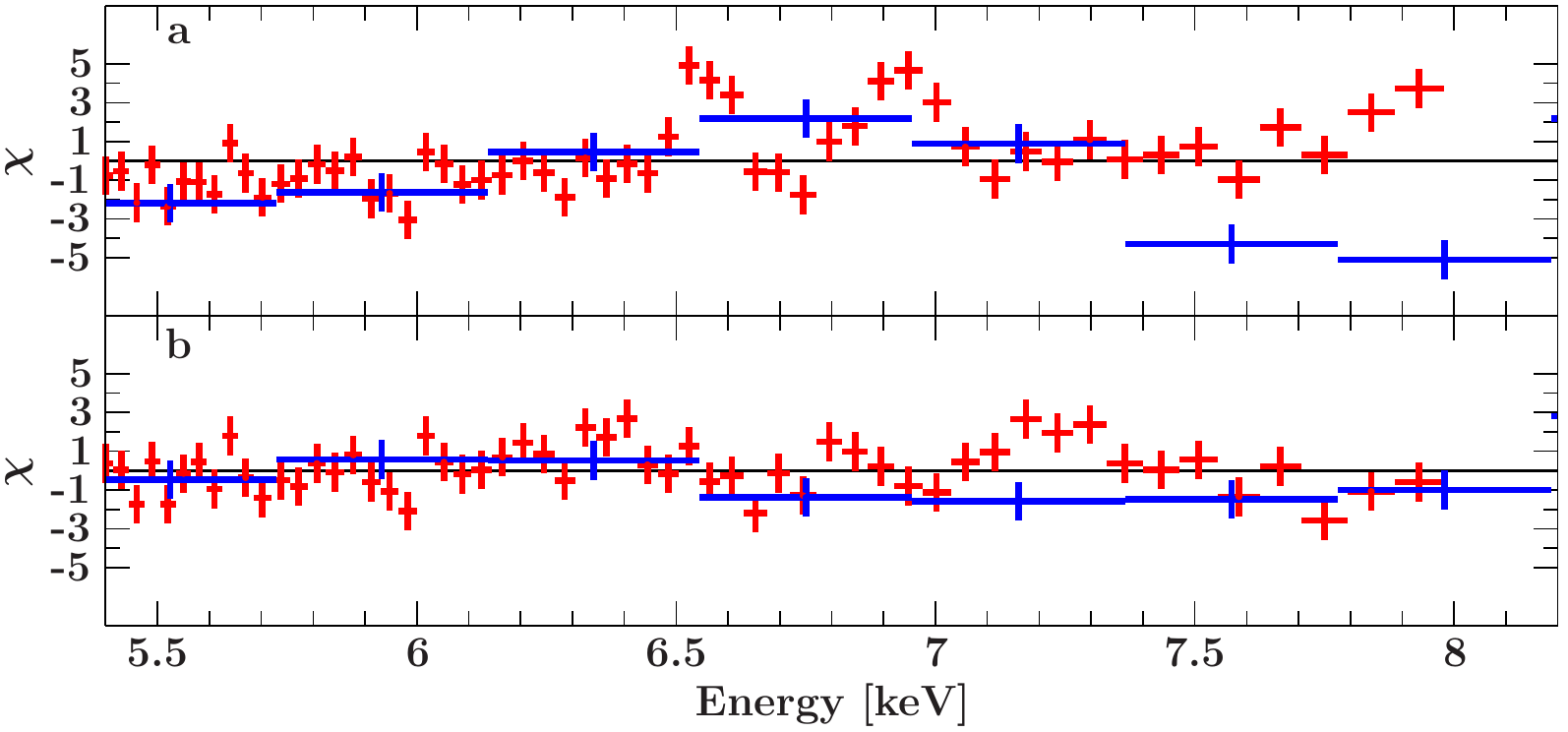}}
  \caption{Test of the gain shift using RXTE data. (a) Fit to the
    \pn(red) and \pca(blue) data with no gain-shift applied. There are
    large residuals, but also a slight energy shift between the
    instruments. (b) Residuals after applying the gain-shift
    correction to \pn data.}
  \label{fig:pcaGain}
\end{figure}

As discussed in Sect.~\ref{sec:gainshift}, in our analysis of Cyg~X-1
we obtained the gain shift parameters from the simultaneous fit of the
\xmm data to the RXTE-PCA, -HEXTE, and INTEGRAL-ISGRI observations.
This method is based on the assumption that the real energy of the
photons, $E_\mathrm{real}$, is linearly related to the energy assigned
by the SAS, $E_\mathrm{obs}$,
\begin{equation}\tag{\ref{eq:gainshift}}
E_\mathrm{real}=E_\mathrm{obs}/s+\Delta E
\end{equation}
Here $s$ is the gain shift parameter and $\Delta E$ is an energy
offset, which is expected to be 0\,keV for CTE effects. 

When performing a standard X-ray data analysis, individual photons are
histogrammed into ``spectra'' with channel boundaries based on
$E_\mathrm{obs}$ rather than $E_\mathrm{real}$, that is, the spectrum
is distorted by the gain shift. In order to correct for the CTE we use
a feature of ISIS that allows us to apply the energy bounds of the
response matrix according to Eq.~\eqref{eq:gainshift}. A similar
feature is also available in XSPEC \citep{arnaud:96a}. Both, $s$ and
$\Delta E$ can be used in spectral fits, that is, by fitting an
observation that is affected by an additional gain shift
simultaneously with observations that are not affected by gain shift,
it is in principle possible to recover $s$ and $\Delta E$. Here we
demonstrate by a detailed comparison with other observations of the
source that this approach yields indeed the correct value for the gain
shift.

We use the \xmm observation Obs2 for this example, as it is the most
stable one in flux and since this observation is comparable in flux
level to an archival \chandra observation of the source, namely
\chandra ObsID~3814 \citep[see][for a thorough analysis of the
  \chandra observation]{hanke:2009}. Fitting the 5.5--8.0\,keV band in
the \chandra spectrum with an absorbed power-law reveals the presence
of a narrow emission line at 6.4 keV and of two absorption lines from
\ion{Fe}{xxv} and \ion{Fe}{xxvi} at $6.689^{+0.010}_{-0.012}$ keV and
$6.98^{+0.12}_{-0.00}$ keV (Fig.~\ref{fig:chandraGain}a). Applying the
best-fit \chandra spectrum to the \pn data leads to strong residuals
around 6.5\,keV (Fig.~\ref{fig:xmmGain}b). The shape of the residuals
is typical for a problem with the instrumental gain. Fixing the
emission and absorption line energies to the values from \chandra,
introducing a gain shift model and refitting improves the fit
statistics to $\chi^{2}_\mathrm{red}=1.79$. The gain-shift slope is
$s_\mathrm{gain}=1.019^{+0.004}_{-0.006}$. Figure~\ref{fig:xmmGain}
shows the significant improvement of the fit. The Fe emission line residuals
have been removed and only small residuals remain at the location of
the absorption lines. Since these occur in the wind and are strongly
variable \citep{hanke:2009,miskovicova:2013}, this is not surprising.

As a cross check we work in the opposite direction and fit a continuum
with absorption and emission lines to the \pn data without using the
gain-shift. The emission line is found at $\mathrm{E}\sim6.56$\,keV,
the absorption lines at $\mathrm{E}\sim6.73$\,keV and
$\mathrm{E}\sim7.0$\,keV. Applying this model to the \chandra data and
only allowing the continuum to change does not result in a good fit.
Figure~\ref{fig:chandraGain}b clearly shows the misplacement,
especially of the emission line (the absorption lines cannot be well
constrained in the \pn spectra).

In order to further strengthen our results from line fitting, we use
the absorption line parameters obtained from the \chandra data and fit
the continuum to the joint \xmm and \rxte data using a broken power
law in the energy range 5.5--8.0\,keV. Figure~\ref{fig:pcaGain} shows
that a simple simultaneous fit leads to $\chi_\mathrm{red}^{2}=4.19$,
which is clearly an unacceptable value and the residuals are again
indicative of a ``best effort'' of the fit to describe two
incompatible spectral shapes (remember that these data were taken
simultaneously). On the other hand, when a gain-shift of
$s_\mathrm{gain}\sim1.024$ is applied on the \pn data, the fit
improves to $\chi_\mathrm{red}^{2}=1.63$ (see Fig.~\ref{fig:pcaGain}).

We conclude, therefore, that continuum fitting and the direct
comparison of the location of distinct features (absorption/emission
lines) gives consistent results and that a gain-shift of $\sim$2\% is
needed in order to describe the \pn spectra. In
Sect.~\ref{sec:relline} we therefore use the values from
Table~\ref{tab:chandra_fit} to describe features that the pn cannot
resolve. For the different observations we also let the gain-shift
slope $s_\mathrm{gainshift}$ free since the above study indicates that
for moderate energy shifts continuum fitting is sensitive enough to
constrain the gain shift.

\onecolumn
\section{Full table values}\label{app:D}

\begin{table}[h]
  \caption{Best fit parameters for a coronal model using a cutoff
    power-law and relativistic reflection with fixed emissivity 
    index $\epsilon=3$ (Model~1). See Sect.~\ref{sec:refl3} for
    more details.} 
  \centering
\renewcommand{\arraystretch}{1.3}
\begin{tabular}{lllll} 
 \hline 
Parameter  & Obs1  & Obs2  & Obs3  & Obs4  \\ 
\hline  
$A_\mathrm{pl}$   &   $1.58^{+0.12}_{-0.10}$   &   $1.10^{+0.06}_{-0.09}$   &   $1.53^{+0.08}_{-0.09}$   &   $1.56\pm0.10$    \\ 
$\Gamma_\mathrm{pl}$   &   $1.724\pm0.012$   &   $1.611^{+0.013}_{-0.016}$   &   $1.609^{+0.012}_{-0.010}$   &   $1.577^{+0.015}_{-0.012}$    \\ 
$E_\mathrm{{fold}}$  [keV]   &   $181\pm14$   &   $175\pm13$   &   $181^{+12}_{-8}$   &   $164^{+9}_{-8}$    \\ 
\hline  
$A_\mathrm{disk}$   &   $\left(161^{+8}_{-7}\right)\times10^{2}$   &   $\left(69\pm7\right)\times10^{2}$   &   $\left(117^{+7}_{-8}\right)\times10^{2}$   &   $\left(147^{+12}_{-13}\right)\times10^{2}$    \\ 
$kT_\mathrm{disk}$  [keV]   &   0.50   &   0.50   &   0.50   &   0.50    \\ 
\hline  
$A_\mathrm{ref}$  [$10^{-5}$]    &   $1.39^{+0.14}_{-0.16}$   &   $1.02^{+0.18}_{-0.13}$   &   $1.33^{+0.20}_{-0.12}$   &   $1.30^{+0.29}_{-0.17}$    \\ 
$\mathrm{Fe}/\mathrm{Fe}_\odot$   &   $3.5^{+0.7}_{-0.6}$   &   $3.1^{+0.8}_{-0.7}$   &   $3.8^{+0.6}_{-0.7}$   &   $3.6^{+1.0}_{-0.7}$    \\ 
$\Gamma$   &   1.72   &   1.61   &   1.61   &   1.58    \\ 
$\xi$  $[\mathrm{erg}\,\mathrm{cm}\,\mathrm{s}^{-1}]$   &   $2330^{+200}_{-190}$   &   $2040^{+250}_{-380}$   &   $2080^{+170}_{-280}$   &   $2100^{+250}_{-410}$    \\ 
\hline  
$\epsilon_{1}$   &   3.0   &   3.0   &   3.0   &   3.0    \\ 
$a$   &   $0.998^{+0.000}_{-0.029}$   &   $0.998^{+0.000}_{-0.060}$   &   $0.998^{+0.000}_{-0.060}$   &   $0.89^{+0.12}_{-0.09}$    \\ 
$i$ [deg]   &   $31.1^{+1.9}_{-1.5}$   &   $27.8^{+3.6}_{-2.4}$   &   $29.8^{+1.7}_{-1.5}$   &   $29.6^{+2.2}_{-3.1}$    \\ 
\hline  
$F_\mathrm{6.4\,keV}$  [$10^3\,\mathrm{cgs}$]   &   $1.44\pm0.20$   &   $0.75\pm0.18$   &   $0.75\pm0.19$   &   $1.52\pm0.29$    \\ 
$E_\mathrm{gauss}$  [keV]   &   6.40   &   6.40   &   6.40   &   6.40    \\ 
\hline  
$E_\mathrm{gabs}$  [keV]   &   6.646   &   6.646   &   6.646   &   6.646    \\ 
$E_\mathrm{\sigma}$  [keV]   &   0.028   &   0.028   &   0.028   &   0.028    \\ 
$\tau$  [$10^{-3}$]    &   $\le0.6$   &   $1.8^{+2.5}_{-1.8}$   &
$4.2^{+2.2}_{-2.0}$   &   $\le 1.82$    \\ 
$E_\mathrm{gabs}$  [keV]   &   6.955   &   6.955   &   6.955   &   6.955    \\ 
$E_\mathrm{\sigma}$  [keV]   &   0.0013   &   0.0013   &   0.0013   &   0.0013    \\ 
$\tau$  [$10^{-03}$]    &   $\le1.2$   &   $\le2.2$   &   $\le1.7$   &   $1.8^{+3.5}_{-1.8}$    \\ 
\hline  
$c_\mathrm{HEXTE}$   &   $0.843\pm0.006$   &   $0.826\pm0.006$   &   $0.834\pm0.004$   &   $0.826\pm0.005$    \\ 
$c_\mathrm{PN}$   &   $0.8377^{+0.0049}_{-0.0029}$   &   $0.780^{+0.008}_{-0.004}$   &   $0.8139^{+0.0026}_{-0.0029}$   &   $0.813\pm0.004$    \\ 
$c_\mathrm{ISGRI}$   &   $0.922^{+0.010}_{-0.009}$   &   $0.921\pm0.009$   &   $0.903\pm0.008$   &   $0.905\pm0.008$    \\ 
\hline  
$s_\mathrm{gainshift}$   &   $1.0266^{+0.0010}_{-0.0023}$   &   $1.0180^{+0.0020}_{-0.0044}$   &   $1.0184^{+0.0012}_{-0.0008}$   &   $1.0209^{+0.0019}_{-0.0012}$    \\ 
\hline  
$\chi^{2}/\mathrm{dof}$ &  574.0/393&  418.7/320&  582.1/443&  406.1/332   \\ 
$\chi^{2}_\mathrm{red}$ &  1.46&  1.31&  1.31&  1.22   \\ 
\hline 
\end{tabular}

\end{table}
  
\begin{table}
  \caption{Best fit parameters for a coronal model using a cutoff
    power-law and relativistic reflection with free emissivity index
    $\epsilon$ (Model~2). See Sect.~\ref{subsec:emmFre} for more
    details.} \centering
  \renewcommand{\arraystretch}{1.3}
\begin{tabular}{lllll} 
 \hline 
Parameter  & Obs1  & Obs2  & Obs3  & Obs4  \\ 
\hline  
$A_\mathrm{pl}$   &   $1.77^{+0.08}_{-0.12}$   &   $1.20^{+0.07}_{-0.08}$   &   $1.57^{+0.07}_{-0.10}$   &   $1.62^{+0.05}_{-0.10}$    \\ 
$\Gamma_\mathrm{pl}$   &   $1.736\pm0.013$   &   $1.630^{+0.017}_{-0.018}$   &   $1.611^{+0.013}_{-0.010}$   &   $1.583^{+0.010}_{-0.013}$    \\ 
$E_\mathrm{{fold}}$  [keV]   &   $172^{+12}_{-10}$   &   $181\pm16$   &   $179^{+12}_{-9}$   &   $163^{+9}_{-5}$    \\ 
\hline  
$A_\mathrm{disk}$   &   $\left(187^{+10}_{-12}\right)\times10^{2}$   &   $\left(62^{+8}_{-7}\right)\times10^{2}$   &   $\left(115\pm8\right)\times10^{2}$   &   $\left(155^{+14}_{-16}\right)\times10^{2}$    \\ 
$kT_\mathrm{disk}$  [keV]   &   0.50   &   0.50   &   0.50   &   0.50    \\ 
\hline  
$A_\mathrm{ref}$  [$10^{-5}$]    &   $1.33^{+0.19}_{-0.13}$   &   $1.6\pm0.4$   &   $1.36^{+0.35}_{-0.14}$   &   $1.30^{+0.20}_{-0.18}$    \\ 
$\mathrm{Fe}/\mathrm{Fe}_\odot$   &   $2.7^{+0.6}_{-0.5}$   &   $3.9\pm0.9$   &   $3.9^{+0.7}_{-0.8}$   &   $3.3^{+0.8}_{-0.6}$    \\ 
$\Gamma$   &   1.74   &   1.63   &   1.61   &   1.58    \\ 
$\xi$  $[\mathrm{erg}\,\mathrm{cm}\,\mathrm{s}^{-1}]$   &   $2030^{+180}_{-310}$   &   $1340^{+610}_{-190}$   &   $2010^{+210}_{-410}$   &   $1930^{+280}_{-260}$    \\ 
\hline  
$\epsilon_{1}$   &   $10.0^{+0.0}_{-1.7}$   &   $3.49^{+0.25}_{-0.31}$   &   $3.08^{+0.19}_{-0.17}$   &   $10^{+0}_{-4}$    \\ 
$a$   &   $-0.59\pm0.20$   &   $0.998^{+0.000}_{-0.028}$   &   $0.998^{+0.000}_{-0.080}$   &   $-0.36^{+0.30}_{-0.24}$    \\ 
$i$ [deg]   &   $33.4^{+1.0}_{-0.8}$   &   $37.6^{+2.5}_{-5.5}$   &   $32^{+4}_{-5}$   &   $33.1^{+1.5}_{-1.4}$    \\ 
\hline  
$F_\mathrm{6.4\,keV}$  [$10^3\,\mathrm{cgs}$]   &   $1.20\pm0.20$   &   $0.83\pm0.19$   &   $0.79^{+0.20}_{-0.22}$   &   $1.34^{+0.32}_{-0.29}$    \\ 
$E_\mathrm{gauss}$  [keV]   &   6.40   &   6.40   &   6.40   &   6.40    \\ 
\hline
$E_\mathrm{gabs}$  [keV]   &   6.646   &   6.646   &   6.646   &   6.646    \\ 
$E_\mathrm{\sigma}$  [keV]   &   0.028   &   0.028   &   0.028   &   0.028    \\ 
$\tau$  [$10^{-3}$]    &   $\le1.2$   &   $1.6^{+2.7}_{-1.6}$   & $3.9^{+2.2}_{-2.0}$   &   $\le +8.23$    \\ 
$E_\mathrm{gabs}$  [keV]   &   6.955   &   6.955   &   6.955   &   6.955    \\ 
$E_\mathrm{\sigma}$  [keV]   &   0.0013   &   0.0013   &   0.0013   &   0.0013    \\ 
$\tau$  [$10^{-3}$]    &   $2.1^{+3.2}_{-2.1}$   &   $\le2.6$   & $\le +1.52$   &   $4^{+5}_{-4}$    \\ 
\hline   
$c_\mathrm{HEXTE}$   &   $0.845^{+0.006}_{-0.005}$   &   $0.825\pm0.006$   &   $0.834\pm0.004$   &   $0.827\pm0.005$    \\ 
$c_\mathrm{PN}$   &   $0.8436^{+0.0057}_{-0.0029}$   &   $0.780\pm0.005$   &   $0.8141^{+0.0026}_{-0.0033}$   &   $0.812\pm0.004$    \\ 
$c_\mathrm{ISGRI}$   &   $0.924\pm0.009$   &   $0.919^{+0.010}_{-0.009}$   &   $0.903\pm0.008$   &   $0.906\pm0.008$    \\ 
\hline  
$s_\mathrm{gainshift}$   &   $1.0205^{+0.0009}_{-0.0025}$   &   $1.0185^{+0.0027}_{-0.0030}$   &   $1.0184^{+0.0013}_{-0.0008}$   &   $1.0204^{+0.0014}_{-0.0018}$    \\ 
\hline  
$\chi^{2}/\mathrm{dof}$ &  541.1/392&  411.0/319&  581.4/442&  391.3/331   \\ 
$\chi^{2}_\mathrm{red}$ &  1.38&  1.29&  1.32&  1.18   \\ 
\hline 
\end{tabular}

\end{table}

\begin{table}
  \caption{Best fit parameters for a coronal model using a cutoff
    power-law and relativistic reflection where the emissivity 
    follows a broken power-law (Model~3). See Sect.~\ref{subsec:broken} for
    more details.} 
  \centering
\renewcommand{\arraystretch}{1.3}
\begin{tabular}{lllll} 
 \hline 
Parameter  & Obs1  & Obs2  & Obs3  & Obs4  \\ 
\hline  
$A_\mathrm{pl}$   &   $1.88\pm0.08$   &   $1.08\pm0.10$   &   $1.60^{+0.05}_{-0.10}$   &   $1.66^{+0.08}_{-0.07}$    \\ 
$\Gamma_\mathrm{pl}$   &   $1.748\pm0.014$   &   $1.600^{+0.022}_{-0.019}$   &   $1.614^{+0.009}_{-0.013}$   &   $1.587^{+0.016}_{-0.015}$    \\ 
$E_\mathrm{{fold}}$  [keV]   &   $177^{+14}_{-13}$   &   $170\pm9$   &   $180^{+10}_{-9}$   &   $164^{+12}_{-10}$    \\ 
\hline  
$A_\mathrm{disk}$   &   $\left(163\pm10\right)\times10^{2}$   &   $\left(47^{+9}_{-27}\right)\times10^{2}$   &   $\left(97^{+20}_{-23}\right)\times10^{2}$   &   $\left(138^{+15}_{-20}\right)\times10^{2}$    \\ 
$kT_\mathrm{disk}$  [keV]   &   0.50   &   0.50   &   0.50   &   0.50    \\
\hline  
$A_\mathrm{ref}$  [$10^{-5}$]    &   $1.83^{+0.30}_{-0.31}$   &   $1.13^{+0.35}_{-0.12}$   &   $1.36^{+0.22}_{-0.17}$   &   $1.6^{+0.5}_{-0.4}$    \\ 
$\mathrm{Fe}/\mathrm{Fe}_\odot$   &   $3.6^{+0.7}_{-0.4}$   &   $6.0^{+0.0}_{-1.7}$   &   $4.4^{+1.6}_{-1.0}$   &   $4.3^{+1.3}_{-0.8}$    \\ 
$\xi$  $[\mathrm{erg}\,\mathrm{cm}\,\mathrm{s}^{-1}]$   &   $1570^{+330}_{-230}$   &   $\left(22^{+4}_{-5}\right)\times10^{2}$   &   $2000^{+310}_{-290}$   &   $\left(17^{+5}_{-4}\right)\times10^{2}$    \\ 
\hline
$\epsilon_{1}$   &   $10.0^{+0.0}_{-3.0}$   &   $5.4^{+4.6}_{-0.7}$   &   $7.5^{+2.5}_{-3.1}$   &   $10^{+0}_{-6}$    \\ 
$\epsilon_{2}$   &   3.0   &   3.0   &   3.0   &   3.0    \\ 
$r_\mathrm{br}$  $[GM/c^2]$   &   $3.38^{+0.27}_{-0.15}$   &   $4.0^{+0.7}_{-0.6}$   &   $3.3^{+0.7}_{-0.4}$   &   $3.31^{+0.78}_{-0.24}$    \\ 
a   &   $0.856^{+0.026}_{-0.020}$   &   $0.989^{+0.009}_{-0.088}$   &   $0.91^{+0.05}_{-0.07}$   &   $0.86\pm0.05$    \\ 
$i$  $[\mathrm{deg}]$   &   $34.1^{+2.4}_{-1.8}$   &   $28\pm4$   &   $30.2^{+1.6}_{-2.5}$   &   $32.0^{+2.8}_{-2.9}$    \\ 
\hline  
$F_\mathrm{6.4\,keV}$  [$10^3\,\mathrm{cgs}$]   &   $1.37\pm0.20$   &   $0.73^{+0.19}_{-0.18}$   &   $0.73\pm0.20$   &   $1.48^{+0.30}_{-0.31}$    \\ 
$E_\mathrm{gauss}$  [keV]   &   6.40   &   6.40   &   6.40   &   6.40    \\ 
\hline 
$E_\mathrm{gabs}$  [keV]   &   6.646   &   6.646   &   6.646   &   6.646    \\ 
$\tau$  [$10^{-3}$]    &   $\le1.2$   &   $3.2\pm2.7$   & $4.7\pm2.0$   &   $\le 4.17$    \\ 
$E_\mathrm{gabs}$  [keV]   &   6.955   &   6.955   &   6.955   &   6.955    \\ 
$\tau$  [$10^{-3}$]    &   $1.6^{+2.8}_{-1.6}$   &   $\le2.0$   & $\le 2.39$   &   $4^{+6}_{-4}$    \\ 
\hline  
$c_\mathrm{HEXTE}$   &   $0.844\pm0.006$   &   $0.823\pm0.007$   &   $0.834\pm0.004$   &   $0.826\pm0.005$    \\ 
$c_\mathrm{PN}$   &   $0.838^{+0.010}_{-0.004}$   &   $0.782^{+0.007}_{-0.004}$   &   $0.8134^{+0.0030}_{-0.0036}$   &   $0.813\pm0.004$    \\ 
$c_\mathrm{ISGRI}$   &   $0.923\pm0.009$   &   $0.916\pm0.010$   &   $0.903\pm0.008$   &   $0.905\pm0.008$    \\ 
\hline  
$s_\mathrm{gainshift}$   &   $1.0251^{+0.0016}_{-0.0010}$   &   $1.0225^{+0.0023}_{-0.0032}$   &   $1.0184^{+0.0015}_{-0.0007}$   &   $1.0213^{+0.0018}_{-0.0015}$    \\ 
\hline  
$\chi^{2}/\mathrm{dof}$ &  544.6/386&  394.6/318&  574.3/441&  397.7/330   \\ 
$\chi^{2}_\mathrm{red}$ &  1.41&  1.24&  1.30&  1.21   \\ 
\hline 
\end{tabular}

\end{table}

\begin{table}
  \caption{Best fit parameters for a lamp post model (Model~4). See 
    Sect.~\ref{subsect:lp} for more details.} 
  \centering
\renewcommand{\arraystretch}{1.3}
\begin{tabular}{lllll} 
 \hline 
Parameter  & Obs1  & Obs2  & Obs3  & Obs4  \\ 
\hline

$A_\mathrm{pl}$   &   $1.91^{+0.10}_{-0.09}$   &   $1.21^{+0.07}_{-0.08}$   &   $1.58^{+0.07}_{-0.06}$   &   $1.67^{+0.10}_{-0.14}$    \\
$\Gamma_\mathrm{pl}$   &   $1.754^{+0.015}_{-0.013}$   &   $1.622^{+0.029}_{-0.017}$   &   $1.614^{+0.014}_{-0.009}$   &   $1.591^{+0.017}_{-0.020}$    \\ 
$E_\mathrm{{fold}}$  [keV]   &   $181^{+15}_{-13}$   &$171^{+29}_{-12}$   &   $180\pm10$   & $166\pm10$  \\ 
\hline  
$A_\mathrm{disk}$   &   $\left(164^{+10}_{-14}\right)\times10^{2}$   &$\left(54^{+10}_{-12}\right)\times10^{2}$   &$\left(116^{+8}_{-9}\right)\times10^{2}$&$\left(147^{+10}_{-13}\right)\times10^{2}$    \\ 
$kT_\mathrm{disk}$  [keV]   &   0.50   &   0.50   &   0.50   &   0.50    \\
\hline  
$A_\mathrm{ref}$  [$10^{-5}$]    &   $2.0\pm0.4$   &   $1.6^{+0.5}_{-0.4}$   &   $1.41^{+0.39}_{-0.15}$   &   $1.7^{+0.6}_{-0.5}$    \\
$\mathrm{Fe}/\mathrm{Fe}_\odot$   &   $3.6^{+0.9}_{-0.6}$   &   $6.0^{+2.5}_{-2.8}$   &   $3.9^{+0.8}_{-0.7}$   &   $3.9^{+1.0}_{-0.8}$ \\
$\xi$  $[\mathrm{erg}\,\mathrm{cm}\,\mathrm{s}^{-1}]$   &   $1440^{+310}_{-200}$   &   $1370^{+410}_{-230}$   &   $1910^{+190}_{-420}$   &   $1600^{+700}_{-400}$    \\
\hline  
$h$   &   $2.58^{+0.37}_{-0.30}$   &   $1.8^{+0.9}_{-0.4}$   &   $3.1\pm0.6$   &   $2.9^{+0.9}_{-0.7}$    \\ 
$a$   &   $0.921^{+0.077}_{-0.027}$   &   $0.995^{+0.004}_{-0.035}$   &   $0.9960^{+0.0020}_{-0.0837}$   &   $0.91^{+0.10}_{-0.07}$    \\ 
$i$ [deg]   &   $38.3^{+1.9}_{-2.0}$   &   $38.2^{+2.2}_{-3.3}$   &   $33.8^{+2.6}_{-2.3}$   &   $36^{+4}_{-6}$    \\ 
\hline
$F_\mathrm{6.4\,keV}$  [$10^3\,\mathrm{cgs}$]   & $1.49\pm0.21$   &   $0.89\pm0.18$   &   $0.83\pm0.20$   &   $1.54^{+0.32}_{-0.29}$    \\
$E_\mathrm{gauss}$  [keV]   &   6.40   &   6.40   &   6.40   &   6.40    \\ 
\hline
$E_\mathrm{gabs}$  [keV]   &   6.646   &   6.646   &   6.646   &   6.646    \\
$\tau$  [$10^{-3}$]    & $\le0.0006$   &   $0.0013^{+0.0024}_{-0.0013}$   &   $0.0035\pm0.0020$   &   $\le0.0021$    \\
$E_\mathrm{gabs}$  [keV]   &   6.955   &   6.955   &   6.955   &   6.955    \\ 
$\tau$  [$10^{-3}$]    &  $\le 0.0029$   &   $\le0.0020$   &   $\le 0.0013$ & $0.0028^{+0.0039}_{-0.0028}$    \\
\hline  
$c_\mathrm{HEXTE}$   &   $0.844\pm0.006$   &   $0.823\pm0.006$   &   $0.834\pm0.004$   &   $0.826\pm0.005$    \\ 
$c_\mathrm{PN}$    &   $0.838^{+0.007}_{-0.005}$   &   $0.779^{+0.004}_{-0.005}$   &   $0.8132^{+0.0025}_{-0.0029}$   &   $0.813^{+0.004}_{-0.005}$    \\ 
$c_\mathrm{ISGRI}$   &   $0.923\pm0.009$   &   $0.916^{+0.010}_{-0.009}$   &   $0.903\pm0.008$   &   $0.905\pm0.008$    \\
\hline  
$s_\mathrm{gainshift}$   &   $1.0253^{+0.0018}_{-0.0029}$   &   $1.0199^{+0.0027}_{-0.0043}$   &   $1.0183^{+0.0010}_{-0.0009}$   &   $1.0210^{+0.0020}_{-0.0012}$    \\ 
\hline  
$\chi^{2}/\mathrm{dof}$ &  574.1/392&  410.4/319&  583.7/442&  405.8/331   \\ 
$\chi^{2}_\mathrm{red}$ &  1.46&  1.29&  1.32&  1.23   \\ 
\hline 
\end{tabular}

\end{table}

\clearpage
\listofobjects
\end{document}